\documentclass[12pt]{iopart}

\usepackage{graphicx}% Include figure files
\usepackage [latin1,utf8]{inputenc}%latin1,
\usepackage{cite}
\usepackage{xcolor}
\usepackage[english]{babel}
\usepackage{hyperref}
\usepackage{dirtytalk}
\usepackage{amsmath}
\usepackage[letterpaper,top=2cm,bottom=2cm,left=2cm,right=2cm,marginparwidth=1.75cm]{geometry}
\makeatletter
\newcommand{\newparallel}{\mathrel{\mathpalette\new@parallel\relax}}
\newcommand{\new@parallel}[2]{%
  \begingroup
  \sbox\z@{$#1T$}% get the height of an uppercase letter
  \resizebox{!}{\ht\z@}{\raisebox{\depth}{$\m@th#1/\mkern-5mu/$}}%
  \endgroup
}
\makeatother
\usepackage{stmaryrd}

\DeclareRobustCommand{\rchi}{{\mathpalette\irchi\relax}}
\newcommand{\irchi}[2]{\raisebox{\depth}{$#1\chi$}} % inner command, used by \rchi
\newcommand{\be}{\begin{equation}}
\newcommand{\ee}{\end{equation}}

\begin{document}

\title[A new quasilinear model for estimating momentum transport]{A new quasilinear model for turbulent momentum transport in tokamaks with flow shear and plasma shaping}

\author[]{Haomin Sun$^{*}$, Justin Ball, Stephan Brunner, Arnas Vol\v{c}okas}
%\author[]{Justin Ball}
%\author[]{Stephan Brunner}

\address{Ecole Polytechnique Fédérale de Lausanne (EPFL), Swiss Plasma Center (SPC), CH-1015 Lausanne, Switzerland}
\ead{$^{*}$haomin.sun@epfl.ch}
\vspace{10pt}
\begin{indented}
\item[]November 2023
\end{indented}

\begin{abstract}\label{abstract}
\textcolor{black}{In tokamak experiments, sufficiently strong $E\times B$ flow shear reduces turbulent transport, thereby improving the prospects for fusion power plants.} It is therefore of great importance to efficiently explore parameter space to find where strong plasma flow can be achieved. To this end, we propose a new, physically motivated quasi-linear model for estimating momentum transport from turbulence in the presence of toroidal flow shear and plasma shaping. The method gives good estimates of momentum transport for up-down asymmetric geometries as well as low magnetic shear and tight aspect ratio. The results are benchmarked with high-fidelity nonlinear GENE simulations, demonstrating that it provides a fast and accurate estimate of momentum transport. 
%, which was understood to be optimal for flow shear suppression of turbulence

\end{abstract}

%\maketitle

\section{Introduction}\label{introduction}%The introduction part of the real paper!
Due to axisymmetry, tokamak plasmas are free to rotate in the toroidal direction, which is composed of parallel (i.e., parallel to the magnetic field) and perpendicular $E\times B$ flow. While this rotation is typically modest ($\sim 10\%$ of the sound speed) \cite{JETJ.M.Noterdaeme_2003,JETdeVries_2006}, faster rotation could be very beneficial. Sufficiently fast rotation can improve MHD stability \cite{RotationMHDBondeson1994PRL,RotationMHDStrait1995PRL,RotationMHDChu1999NF,RotationMHDWahlberg2000POP,RotationMHDGarofalo2002PRL,RotationMHDAiba2009NF,RotationMHDAiba2011NF} and gradients in rotation (in particular, associated with $E\times B$ flow shear) can reduce turbulent transport \cite{Stambaugh1990enhanced,Biglari1990,JETL-GEriksson_1997,JETMantica2009PRL,Ida2009PRL,JETdeVries_2008,Angioni2001IntrinsicRotation,schekochihin2008,NewtonFlowShearUnderstanding2010,barnes2011,BarnesFlowShear2011,highcock2012,schekochihin2012,ChristenFlowShear2018,ben2019}. On the other hand, the radial gradient of the parallel flow can strengthen turbulence \cite{Peeters2005LinearToroidal,schekochihin2008,BarnesFlowShear2011,schekochihin2012,ball2019} by destabilizing the so-called Parallel Velocity Gradient (PVG) driven modes. Fortunately, this typically occurs for extreme values of flow shear, \textcolor{black}{well beyond what is needed for $E\times B$ flow shear stabilization and what current tokamak experiments typically achieve\cite{Highcock2011POP,highcock2012zero}}. 

To drive strong plasma rotation, one can use Neutral Beam Injection (NBI) \cite{Groebner1990NBIrotation,Suckewer1981NBIrotation,Goumiri2016NSTXNBI} or Radio Frequency (RF) waves \cite{Hsuan1996ICHrotation,Chang1999ICRH,Chan2002RFrotation,Li_2011EASTRF,Lyu2020RFrotation} to apply an external torque to the plasma. However, external injection is not expected to scale well to large devices \cite{YueqiangLiu_2004}. An attractive alternative is intrinsic rotation, which is rotation generated under certain conditions by turbulence in the plasma. This method has the potential to scale well, as it does not rely on external sources. However, due to the symmetry properties of gyrokinetics \cite{Peeters2005LinearToroidal,ParraUpDownSym2011}, intrinsic rotation is constrained to be slow compared to the sound speed unless the up-down symmetry of the flux surface shapes is broken \cite{ball2014,ball2018}. 

In a steady-state tokamak, these intrinsic drive mechanisms will be balanced by the diffusive turbulent and neoclassical momentum transport \cite{Newton2006neoclassical,Wang2009PRL,Stacey_2014DIIIDneoclassical,Stoltzfus-Dueck_neoclass_2019}, where this diffusive transport is the viscous momentum transport driven by toroidal flow shear \cite{Peeters2005LinearToroidal,Hahm2007NLtheory,Diamond2008momentumtransport,Holod2008GKSim,CassonExBshear2009,Yoon_2010momentumtransportITG,ParraUpDownSym2011,Camenen_2011NFReview,Peeters_2011NFReview,Angioni_2012NFReview,Diamond_2013NFreview,ball2014}. \textcolor{black}{In this paper, we focus on diffusion associated with turbulence as it is typically much more important than neoclassical diffusion \cite{Peeters_2011NFReview,Parratheory_2015,Zimmermann_2022}}. The strength of diffusive processes is quantified by the ion turbulent momentum diffusivity $D_{\Pi_i}$ and is frequently compared against the ion turbulent heat diffusivity $D_{Q_i}$ using the ion Prandtl number \cite{ball2014}
\begin{eqnarray}
Pr_i = \frac{D_{\Pi_i}}{D_{Q_i}}.%=-\frac{\Pi_{i,norm}}{Q_{i,norm}}\frac{c_s}{R_0T_i}\frac{dT_i}{dx}\frac{1}{\frac{\partial|\Omega_{tor}|}{\partial x}}
%\frac{\Pi_{i,norm}}{Q_{i,norm}}\frac{omt}{ExBrate|q_0|}\epsilon
\label{eq_prandtl}
\end{eqnarray}
This ratio of the two diffusivities will be a focus of this work and is important because we want to identify optimal conditions for rotation to reduce energy transport. Indeed, a lower Prandtl number (i.e. a lower momentum transport) means that a given source of momentum (external or intrinsic) will drive stronger rotation shear for a given level of turbulence. Thus, the rotation shear will be more capable of combating turbulence, which is desirable for future tokamak operations \cite{HighcockRotationBifurcation2010,highcock2012,Ben_2019}. Recently, it has been found that a low momentum diffusivity regime can be reached at tight aspect ratio and low safety factor \cite{mcmillan2019}. On the other hand, the efficiency of using flow shear to stabilize turbulence can also depend significantly on other geometric factors. In particular, previous experimental \cite{JETMantica2009PRL,JETMantica2011PRL} and theoretical \cite{BarnesFlowShear2011,highcock2012} works suggest that a combination of flow shear and low magnetic shear can facilitate the stabilization of turbulence. Therefore, in order to find the tokamak configurations that exhibit flow shear suppression of turbulence, we are particularly interested in regimes with strong intrinsic flow drive (i.e. up-down asymmetric shaping), low momentum diffusivity (i.e. tight aspect ratio, low safety factor) and an efficient suppression of turbulence by the flow shear (i.e. low magnetic shear). 

In the literature, there have been many works modelling momentum and heat transport by nonlinear (NL) gyrokinetic simulations \cite{Holod2008GKSim,casson2009,Camenen2009TUBTrans,Camenen2010MomentumTransport,BarnesFlowShear2011,Angioni_2011GKModel,highcock2012}. However, such simulations are computationally expensive, making it costly to explore a large parameter space. In order to save computational time, the so-called Quasi-Linear (QL) models for turbulent transport \cite{QuaLiKizStephens2021,Dudding2022QLsaturation} have been developed. A QL model is a method for estimating NL turbulent fluxes based on linear simulation results. It is typically constructed by considering the normalized contributions of linear fluxes from different eigenmodes and then combining them using QL weightings. Different QL models differ mainly by how they calculate QL estimates of fluxes from different eigenmodes and their QL weights, as well as the number of modes being considered in the model. One prominent example is QuaLiKiz, a well-developed and thoroughly benchmarked QL model for fast modelling of turbulent transport \cite{QuaLiKizC.Bourdelle_2002,QuaLiKizBourdelle2007,QuaLiKizCasati2009,Cottier2014QuaLiKizmomentum,QuaLiKizBaiocchi2015,QuaLiKizCitrin2015,QuaLiKizBourdelle2016,QuaLiKizCitrin2017,QuaLiKizFelici2018,QuaLiKizCasson2020,QuaLiKizPlassche2020,QuaLiKizMarin2021,QuaLiKizStephens2021}. It not only models turbulent transport at low magnetic shear \cite{Citrin2012QuaLiKizmagneticshear}, but also momentum transport due to externally imposed flow shear \cite{Cottier2014QuaLiKizmomentum}. The model has a computational cost that is typically two orders of magnitude lower than NL simulations \cite{QuaLiKizFelici2018}. Despite the success of QuaLiKiz, it relies on several assumptions that limit its applicability. In particular, it assumes circular flux surfaces and a large aspect ratio. The first prevents modeling intrinsic rotation driven by up-down asymmetry. As a consequence of the second assumption, the toroidal angular momentum flux in the model is simplified to the parallel momentum flux. This approximation breaks down at the tight aspect ratio of spherical tokamaks. These configurations are, however, of particular interest to us, as the momentum diffusivity is lower at tight aspect ratio \cite{Ben_2019}. Additionally, while QuaLiKiz can model momentum transport, this particular functionality is only benchmarked at normal values of magnetic shear, whereas we are interested in a combination of flow shear and low magnetic shear \cite{JETMantica2009PRL,JETMantica2011PRL,BarnesFlowShear2011,highcock2012}. As we will see, such a combination is challenging to model because the linear eigenmodes are pushed away from the outboard midplane by the flow shear \cite{QuaLiKizStephens2021}, requiring many ballooning angles to be considered in a proper manner in a QL model. Another successful QL model, known as TGLF, has been \textcolor{black}{widely used by many people} \cite{SATNordman1990,SATStaebler2013,SATStaebler2016,SATStaebler2021,Dudding2022QLsaturation}. The authors of this model carefully examined a database of NL gyrokinetic simulations and fit the QL model. Although it is not a pure first-principles model, it achieves an excellent agreement with NL simulations. There are also other QL models in the literature, such as for Electron Temperature Gradient (ETG) driven turbulence in the pedestal \cite{HatchQLETG}, for stellarators \cite{PueschelQLStellarator} and for microtearing turbulence \cite{XieQLMTturbulence}. However, none of these models, to the best of our knowledge, include an estimate of toroidal angular momentum flux. It is therefore desirable to develop a new QL model that can estimate the toroidal angular momentum transport for different aspect ratios, magnetic shear, flow shear, and plasma shaping (including up-down asymmetry).

In this work, we develop such a new QL model by combining linear gyrokinetic flux tube simulations with the GENE code \cite{FJenko_2001GENEcode,GoerlerGENE2011} and a physically motivated method for estimating QL weights. We simplify our task by only seeking to calculate the ratio of the toroidal angular momentum flux to the heat flux. This is the relevant quantity for estimating the importance of flow shear stabilization of turbulence. Additionally, it can be used to calculate the toroidal angular momentum if one calculates the heat flux using a standard QL code like QuaLiKiz or TGLF. In Sec. \ref{section2}, we first present a basic QL model to familiarize the reader. Then we take the example of intrinsic momentum transport in up-down asymmetric geometries to show how such a basic QL model fails. By extending the model to include multiple ballooning angles, we are able to achieve good agreement with NL simulations. In Sec. \ref{section3}, we consider more complicated cases by accounting for flow shear as well. To this end, we generalize the previous QL model by analyzing the Floquet-type evolution of independent ballooning modes. By following the time-dependent linear growth of the individual ballooning modes, we motivate a natural generalization of our QL model for non-zero flow shear. The resulting QL estimates are then benchmarked against NL GENE simulations, with reasonable agreement for up-down asymmetric geometries, low magnetic shear, and flow shear. The conclusions and discussions of our model, as well as possible experimental applications, are given in Sec. \ref{section4}. The method paves a new way to estimate momentum transport for turbulence in tokamaks with rotational flow shear and plasma shaping. To the best of our knowledge, this is also the first QL model that accurately models \emph{toroidal} angular momentum flux.

\par
%The paper is organized as follows: The invalidity of traditional QL model for estimating flux ratio in up-down asymmetric geometry with normal aspect ratio and our extension to the traditional model are shown in Sec. \ref{section2}. Our newly developed QL model for cases with flow shear and different geometries is presented in Sec. \ref{section3}, where both tight aspect ratio cases with different magnetic shear and more advanced cases are benchmarked with NL GENE simulations. Conclusions and Discussions of our model and the possible experimental applications will be given in Sec. \ref{section4}.

\section{QL estimates of momentum transport in up-down asymmetric geometries}\label{section2}
In this section, we develop a QL model for momentum transport driven by up-down asymmetry in the magnetic geometry. We start from a basic QL model similar to the one in Ref. \cite{Mariani2017}, and show the importance of considering multiple ballooning angles. The results are benchmarked with NL GENE simulations, showing good agreement.
\subsection{Description of the basic and multi-$\rchi_0$ QL model}\label{QLmodelsmall}
In this paper, all the equations and simulations will use the GENE coordinate system \cite{GoerlerGENE2011,Grler2010MultiscaleEI}, which considers the $(x,y,z)$ spatial coordinates and the $(v_{||},\mu)$ velocity coordinates. Here $(x,y,z)$ are the radial, binormal, and straight field line poloidal angle $\rchi$, respectively. As $\vec{B}$ is parallel to $\nabla x\times\nabla y$, $x=const$ and $y=const$ define a magnetic line parameterized by $z$, which is therefore also called the parallel (to the magnetic field) coordinate. In the Fourier space representation used by GENE, the coordinates become $(k_x,k_y,z)$, where $k_x$ and $k_y$ are radial and binormal wave numbers, respectively. The velocity coordinates are the parallel velocity $v_{||}$ and the magnetic moment $\mu=mv_{\perp}^2/2B$, where $m$ is the particle mass, $v_{\perp}$ is the perpendicular velocity, and $B$ is the magnetic field strength. To understand the functional form of a QL estimate, we will start by recalling the structure of a linear eigenmode in the gyrokinetic simulations. As a result of the assumed axisymmetry of the equilibrium state, eigenmodes have a fixed toroidal wave number $n$ corresponding to a given binormal wave number ($k_y=nq_0/r_0$), where $q_0$ and $r_0$ are the safety factor and the minor radius respectively, evaluated at the center of flux tube. As long as the magnetic shear $\hat{s}$ is finite, the parallel boundary condition (along $z$) leads to a linear coupling of a subset of $k_x$ modes: $k_x=k_{x0}+p2\pi k_y\hat{s}$, where $p$ is an integer \cite{Ajay_2020,BeerBallooingCoordinates1995}. Therefore, each linear eigenmode can be characterized by a fixed $k_y$ and a \say{central} radial wavenumber $k_{x0}$ (typically the smallest in absolute value among the coupled subset). Such a linear eigenmode is conveniently represented in the so-called ballooning representation in which the parallel coordinate $z$ is extended to the coordinate $z_b$ in the infinite ballooning space $z_b\in(-\infty,+\infty)$ \cite{Connor1978PRLBallooning,Hazeltine1990POFBallooning}. The transformation between the ballooning representation $\phi_b(\rchi_{0},k_y,z_b)$ and the usual Fourier $\phi(k_x,k_y,z)$ modes for any scalar field (using the electrostatic potential $\phi$ as an example) is
\be\label{eq_Coordinate1}
% \phi_b(\rchi_{0},k_y,z_b,t)=\phi(\rchi=\rchi_0+2\pi k_y\hat{s}P(z_b),k_y,z=z_b-2\pi P(z_b),t), 
\phi_b(\rchi_{0},k_y,z_b)=\phi(k_x=-k_{y}\hat{s}\rchi_0+2\pi k_y\hat{s}P(z_b),k_y,z=z_b-2\pi P(z_b)), 
\ee
%, i.e. along the line with fixed straight field line poloidal angle $\rchi=\rchi_0$ (though it is only an estimate as it does not account for the local magnetic shear).
where $P(z_b)=\text{NINT}(z_b/2\pi)$ and $\text{NINT}(\xi)$ provides the nearest integer to any scalar $\xi$. Other physical quantities defined in $(k_x,k_y)$ Fourier space can also be transformed to their ballooning representation using the same relation as Eq. \ref{eq_Coordinate1}. One also defines the ballooning angle $\rchi_0=-k_{x0}/k_y\hat{s}$, which estimates the straight field line poloidal angle at which the perturbation has wavefronts aligned with the minor radial direction. Different Fourier modes are thus 
\say{connected} in ballooning space to form a single linear mode. For a given $k_y$, one can choose the number of independent \say{ballooning modes} (i.e. the number of independent values of $\rchi_{0}\in (-\pi,\pi]$) considered in a numerical estimate, which will be denoted by 
$M=\text{NINT}(2\pi k_y\hat{s}/\Delta k_x)$ \cite{Dominski2015POP}, where $\Delta k_x$ is the grid spacing in $k_x$. For the construction of QL models, previous works often assumed $M=1$ in order to consider just one linear ballooning mode, typically with $\rchi_{0}=0$, for each $k_y$ \cite{Camenen2009QLtransport,Citrin2012QuaLiKizmagneticshear,Mariani2017,Hornsby_2017}. This approach is often appropriate given that without symmetry breaking effects (such as up-down asymmetry, background shear flow, and profile shearing), the fastest growing linear mode is usually the one with $\rchi_0=0$. This is because it is centered at the outboard midplane and thus maximizes the curvature drive. The basic QL model below takes just such an approach. It estimates NL fluxes according to \cite{Fable2010QLmodel,Mariani2017}
%important aspect of this model is to clarify that the summation range of $k_x$ in Eq. \ref{eq_QL1-2-1}, Eq. \ref{eq_QL1-2-2} and Eq. \ref{eq_QL1-4} is over only one ballooning angle centered at the outboard midplane, which is the same as many previous QL estimates \cite{Camenen2009QLtransport,Citrin2012QuaLiKizmagneticshear,Mariani2017}.
\be\label{eq_QL1-1}
F^{QL} = A_0\sum_{k_y}w^{QL}(\rchi_0=0,k_y)F^{L}_{norm}(\rchi_0=0,k_y),
\ee
where $F$ refers to either the particle flux $\Gamma$, angular momentum flux $\Pi$ or heat flux $Q$, $A_0$ is an overall normalization constant and $w^{QL}$ is the so-called QL weighting of each linear flux, $F^{L}_{norm}$. Importantly, in this paper, we are only interested in the \emph{ratio} between the fluxes (primarily the toroidal angular momentum flux divided by the heat flux), so $A_0$ cancels. The normalized linear flux $F^{L}_{norm}$ for each eigenmode is defined according to

\be\label{eq_QL1-2}
F^{L}_{norm}(\rchi_0,k_y)=\frac{\langle F^{L}_b(\rchi_0,k_y,z_b,t_{\infty})\rangle_{z_b}}{\text{MAX}_{z_b}(|\phi_b(\rchi_0,k_y,z_b,t_{\infty})|^2)},
\ee
%$F^L(k_y,t_{\infty})=\sum_{k_x}\langle F^L(k_x,k_y,z,t_{\infty})\rangle_z$
%in $F^{L}_b$, $F$ can be toroidal angular momentum flux $\Pi=\Pi_{||}+\Pi_{\perp}$ or heat flux $Q$
where $\text{MAX}_{z_b}(...)$ returns the maximum value over $z_b$ and $t_{\infty}$ refers to the final timestep of the linear simulation. Note that one should run the simulation for long enough to achieve convergence. The average over $z_b$ in Eq. \ref{eq_QL1-2} is taken to be $\langle A\rangle_{z_b}=\int^{\infty}_{-\infty} A(z_b)J_b(z_b)dz_b/\int^{\infty}_{-\infty} J_b(z_b)dz_b$, for any arbitrary function $A$. The integral is taken over the entire length of the ballooning mode and $J_b(z_b)=J(z=z_b-2\pi P(z_b))$ is the periodic extension of the coordinate system Jacobian, $J=[(\nabla x\times\nabla y)\cdot\nabla z]^{-1}$. $F^{L}_b$ is the linear flux in ballooning space, which is transformed from the original linear GENE output $F^L(k_x,k_y,z,t_{\infty})$ according to Eq. \ref{eq_Coordinate1}. A \textcolor{black}{general form} of the explicit expression of $F^L(k_x,k_y,z)$ when taking $F$ as $\Gamma$, $\Pi_{||}$, $\Pi_{\perp}$ (where $\Pi=\Pi_{||}+\Pi_{\perp}$), and $Q$ is \cite{ParraUpDownSym2011}
\textcolor{black}{
\be\label{eq_QL1-2-1}
\Gamma_s^{L}(k_x,k_y,z,t)= C\int d^3v h_s\left(\Vec{\boldsymbol{v}}\cdot\nabla x\right) ,
\ee %\right\rangle_{x} \right\rangle_{\Delta x} \right\rangle_{\Delta t}
\be\label{eq_QL1-2-2}
\Pi_{s,||}^{L}(k_x,k_y,z,t)= m_s R C\int d^3v h_s \left(\Vec{\boldsymbol{v}}\cdot\nabla x\right)v_{||}\hat{\boldsymbol{b}}\cdot\hat{\boldsymbol{e}}_{\zeta},
% \Pi^{L}(k_x,k_y,z)=\frac{2}{C}Re\left[ik_y\phi^*\int mv_{\zeta}R\delta fd^3v \right],
\ee
\be\label{eq_QL1-2-2-2}
\Pi_{s,\perp}^{L}(k_x,k_y,z,t)= m_s R C\int d^3v h_s \left(\Vec{\boldsymbol{v}}\cdot\nabla x\right)\Vec{\boldsymbol{v}}_{\perp}\cdot\hat{\boldsymbol{e}}_{\zeta},
\ee
and
\be\label{eq_QL1-2-3}
Q_s^{L}(k_x,k_y,z,t)= \frac{m_s}{2}C \int d^3v h_s \left(\Vec{\boldsymbol{v}}\cdot\nabla x\right)v^2,
\ee}where $h$ is the $x-y$ Fourier transform of the fluctuating part of the particle distribution function, subscripts \say{$s$} denote different particle species, $\zeta$ is the toroidal angle, $\hat{\boldsymbol{e}}_{\zeta}$ is the unit vector in the toroidal direction, $\hat{\boldsymbol{b}}$ is the unit vector along the magnetic field, \textcolor{black}{$C$ is a geometrical coefficient.} A more \textcolor{black}{detailed} version of these expressions are given in \ref{AppendixD}, which is what is actually calculated from the GENE simulations. Inspired by previous work \cite{Mariani2017}, the QL weights are chosen to be 
%, where it is shown explicitly how the toroidal angular momentum flux can be divided into a parallel component $\Pi_{||}$ and a perpendicular component $\Pi_{\perp}$
%Note both $F^L_b$ and $\phi_b$ are not functions of $\rchi_0$ because the basic QL model only includes a single ballooning mode with $\rchi_0=0$.

% The expression of $F^{L}$ for toroidal angular momentum flux and heat flux are\cite{Mariani2017,ball2014}:
% \be\label{eq_QL1-2-1}
% \Pi^{L}(k_y,t_{\infty})=\left\langle\frac{1}{C}\sum_{k_x}2Re\left[ik_y\phi^*\int mv_{\zeta}R\delta fd^3v \right]\right\rangle_z
% \ee
% \be\label{eq_QL1-2-2}
% Q^{L}(k_y,t_{\infty})=\left\langle\frac{1}{C}\sum_{k_x}2Re\left[ik_y\phi^*\int\frac{1}{2}mv^2\delta fd^3v\right]\right\rangle_z
% \ee
% where $\Pi$ can be decomposed into a parallel component and a perpendicular component: $\Pi=\Pi_{||}+\Pi_{\perp}$ \cite{ball2016a}, $\zeta$ is the toroidal angle, $C=B_0/\sqrt{g^{xx}g^{yy}-(g^{xy})^2}$, $B_0$ is background magnetic field, ${g^{ij}=\nabla i\cdot\nabla j, i,j=x,y,z}$. $\langle A\rangle_z=\int A(z)J(z)dz/\int J(z)dz$ is the flux surface average. $J=[(\nabla x\times\nabla y)\cdot\nabla z]^{-1}$ is the Jacobian of $(x,y,z)$ coordinate system. Please note that the summation of $k_x$ is over the linearly connected ballooning mode. Because we take $M=1$, the summation is over one single ballooning mode centered at $\rchi_0=0$ for each $k_y$. 
% The expression of the QL weighting is: 
\be\label{eq_QL1-3}
w^{QL}(\rchi_0,k_y) = 
\left\{
\begin{aligned}
\left(\frac{\gamma(\rchi_0,k_y)}{\langle k^2_{\perp b}\rangle(\rchi_0,k_y)}\right)^{\xi} &\quad\text{if}\quad \gamma(\rchi_0,k_y)>0\\
0 &\quad \text{else},\\%\quad\text{if}\quad\gamma(k_y,t_{\infty})<0
\end{aligned}
\right.
\ee
where $\xi$ is an undetermined exponent. Here, $\gamma(\rchi_0,k_y)$ is the growth rate of the linear eigenmode, which is directly provided by GENE linear flux tube simulations. Unless explicitly noted, all cases in this paper consider $
\xi=4$ because it gives the best agreement when benchmarking the model to NL simulations (as will be shown in Figs. \ref{spectrumnexc=1} and \ref{spectrumnexc=8}). In Eq. \ref{eq_QL1-3}, we must also estimate the average perpendicular wavenumber, which is done by weighting the mode amplitude $|\phi_b(\rchi_0,k_y,z_b,t_{\infty})|^2$ according to
\be\label{eq_QL1-4}
%\langle k^2_{\perp}\rangle(k_y,t_{\infty})=\frac{\sum_{k_x}\int k^2_{\perp}(k_x,k_y,z,t_{\infty})|\phi(k_x,k_y,z,t_{\infty})|^2J(z)dz}{\sum_{k_x} \int |\phi(k_x,k_y,z,t_{\infty})|^2J(z)dz},
\langle k^2_{\perp b}\rangle(\rchi_0,k_y)=\frac{\int^{\infty}_{-\infty} k^2_{\perp b}(\rchi_0,k_y,z_b)|\phi_b(\rchi_0,k_y,z_b,t_{\infty})|^2J_b(z_b)dz_b}{\int^{\infty}_{-\infty} |\phi_b(\rchi_0,k_y,z_b,t_{\infty})|^2J_b(z_b)dz_b}.
\ee
%g^{xx}(z)k^2_x+2g^{xy}(z)k_xk_y+g^{yy}(z)k^2_y
Here $k^2_{\perp b}(\rchi_0,k_y,z_b)$ is obtained by extending $k^2_{\perp}(k_x,k_y,z)=k_x^2|\Vec{\nabla} x|^2+2k_xk_y\Vec{\nabla}x\cdot\Vec{\nabla}y+k_y^2|\Vec{\nabla}y|^2$ into ballooning space using Eq. \ref{eq_Coordinate1}. \par 
%Note that here since we have set $M=1$, all the $k_x$ Fourier modes are part of the same linear mode and share the same linear growth rate.\par
%One important aspect of this model is to clarify that the summation range of $k_x$ in Eq. \ref{eq_QL1-2-1}, Eq. \ref{eq_QL1-2-2} and Eq. \ref{eq_QL1-4} is over only one ballooning angle centered at the outboard midplane, which is the same as many previous QL estimates \cite{Camenen2009QLtransport,Citrin2012QuaLiKizmagneticshear,Mariani2017}. To be specific, in gyrokinetic simulations, we know that each mode is extended in the ballooning coordinate $z_b$ as a "ballooning mode" $\phi_b(k_{x0},k_y,z_b,t)=\phi(k_x=k_{x0}+2\pi k_y\hat{s}P(z_b),k_y,z=z_b-2\pi P(z_b),t)$, where $k_{x0}$ is the center of ballooning mode, $\hat{s}$ is the magnetic shear, $P(z_b)=INT(z_b/2\pi)$. Therefore, the different Fourier modes are actually "connected" in the ballooning space. The number of independent "ballooning modes" (the number of $k_{x0}$) is often denoted by "$M$" \cite{Dominski2015POP}. Previous work often assume $M=1$ and take $k_{x0}=0$. The model above takes the same approach, which is why the $k_x$ can be any value in $\phi$ of Eq. \ref{eq_QL1-5} because different $k_x$ modes are linearly coupled and share the same growth rate. 
Equations \ref{eq_QL1-1} to \ref{eq_QL1-4} define what we will call the \say{basic} QL model, which can successfully estimate the NL fluxes for many cases of interest. However, when the flux surfaces are up-down asymmetric or the magnetic shear $\hat{s}$ is low, such a model will break down. In the first case, the flux surfaces no longer possess symmetry about the midplane, so there is no reason to expect the $\rchi_0=0$ ballooning mode to be the most unstable. In the second case, the mode instability becomes less sensitive to $\rchi_0$, so many values of $\rchi_0$ contribute significantly to the turbulent transport. In such cases, we must therefore consider multiple $\rchi_0$ ballooning angles. Fortunately, there is a straightforward and natural way to do this. We modify the above basic QL model to be
\be\label{eq_QL1-6}
F^{QL} = A_0\sum_{\rchi_{0},k_y}w^{QL}(\rchi_{0},k_y)F^{L}_{norm}(\rchi_{0},k_y).
\ee
\textcolor{black}{A similar approach, but without transforming to ballooning space, is used in TGLF \cite{SATStaebler2021,Dudding2022QLsaturation}.} We see that compared to Eq. \ref{eq_QL1-1}, we have simply added a summation over multiple independent ballooning modes, parameterized by $\rchi_{0}$, each with their own individual weight. The normalized fluxes $F^{L}_{norm}(\rchi_0,k_y)$ for the different ballooning angles are still estimated according to Eq. \ref{eq_QL1-2}.
% \be\label{eq_QL1-7}
% F^{L}_{norm}(\rchi_{0},k_y,t_{\infty})=\frac{\langle F^{L}_b(\rchi_{0},k_y,z_b,t_{\infty})\rangle_{z_b}}{\text{MAX}_{z_b}(|\phi_b({\rchi_{0}},k_y,z_b,t_{\infty})|)^2}.
% %,k_{x}(\rchi_0)
% \ee
In practice, a scan of independent GENE linear simulations is performed for each value of $\rchi_{0}$ and $k_y$. For a given linear simulation, we thus set $M=1$, so the $z_b$ in the expressions represents the ballooning space for an individual $\rchi_0$. The linear fluxes $\Pi^L_b(\rchi_0,k_y,z_b,t_{\infty})$ and $Q^L_b(\rchi_0,k_y,z_b,t_{\infty})$ are again obtained by transforming to ballooning space using Eq. \ref{eq_Coordinate1} from Eqs. \ref{eq_QL1-2-2} to \ref{eq_QL1-2-3}. The QL weighting function $w^{QL}(\rchi_0,k_y)$ is also still estimated according to Eq. \ref{eq_QL1-3}, where we also take $\xi=4$ and the expression for $\langle k^2_{\perp b}\rangle$ remains the same as Eq. \ref{eq_QL1-4}. Similarly, the linear growth rate $\gamma(\rchi_0,k_y)$ is obtained directly from GENE linear simulations. We see that this new QL model composed of Eq. \ref{eq_QL1-6} together with Eqs. \ref{eq_QL1-2} to \ref{eq_QL1-4} is very similar to the \say{basic} QL model of Eqs. \ref{eq_QL1-1} to \ref{eq_QL1-4}. We simply include multiple ballooning angles for each value of $k_y$. We therefore will call this model the \say{multi-$\rchi_0$} QL model for the rest of this paper.

\subsection{QL Model benchmarking with up-down asymmetric nonlinear simulations}\label{updownbenchmark}
\par
   %GENE code is a well-developed and thoroughly benchmarked gyrokinetic simulation code \cite{FJenko_2001GENEcode}. With a recent development dealing with flow shear calculation in local flux-tube simulations \cite{ben2019}, the flux-tube model can perform linear and NL simulations with both positive and negative flow shear. 
   In this section, we will test the basic and multi-$\rchi_0$ QL models for up-down asymmetric cases by comparing them to corresponding standard NL GENE simulations. Table \ref{updownpara} in \ref{AppendixA} gives the grid parameters for both GENE linear and NL simulations. Two sets of linear simulation scans are performed. In the first set, we include only $\rchi_0=0$ with $M=1$ in order to calculate the basic QL model. The second set of linear simulations scans many values of $\rchi_0$ in order to calculate the multi-$\rchi_0$ QL model. Both of these sets of simulations include multiple $k_y$ modes on an equidistant mesh with spacing $\Delta k_y$ to capture the important contributions in the corresponding NL grid. We only consider $k_y$ values up to $k_y\rho_i=1$ as the contribution from larger $k_y$ modes to the momentum and heat flux is negligible in NL simulations. 
   
\begin{figure}
\centering
\includegraphics[width=0.52\textwidth]{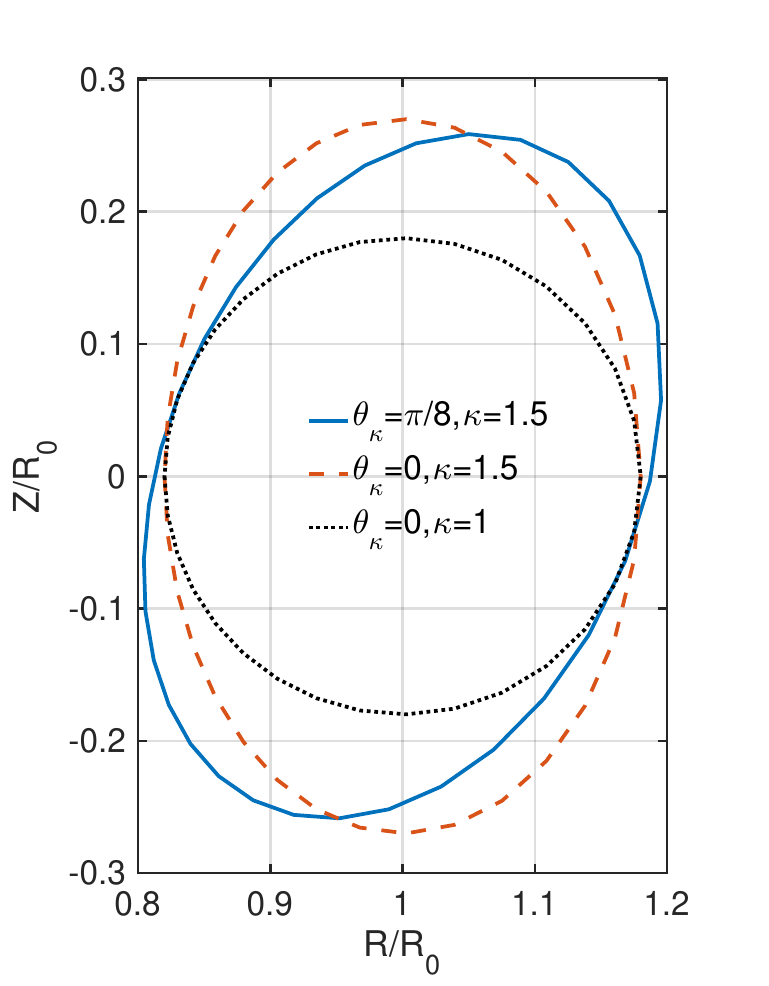}
\caption{\label{geometryupdown} (color online) The circular (black) and tilted elongated (blue) flux surface geometries considered in this paper. The elongated but untilted (red) flux surface is also shown as a reference. $R/R_0$ is the normalized major radial coordinate and $Z/R_0$ is the normalized vertical coordinate.}
%\caption{\label{geometryupdown} Geometric factors (a) Poloidal cut of magnetic surface, (b) $g_{xx}$ as a function of $z$, (c) $g_{xy}$ as a function of $z$, (d) $g_{yy}$ as a function of $z$ for up-down asymmetric geometry with $q=3.05$, $\hat{s}=0.8$, $\epsilon=0.18$, $\rchi_{\kappa}=\pi/8$, $\kappa=1.5$. The geometric factors of $\rchi_{\kappa}=0,\kappa=1.5$ and $\rchi_{\kappa}=0,\kappa=1.0$ are also plotted as a reference.}
\end{figure}

   Table \ref{updownphysicspara} gives the physical parameters of the simulations. Here we consider flux surfaces with an aspect ratio $\epsilon=0.18$ and elongation $\kappa=1.5$ but with the elongation tilted by an angle $\theta_{\kappa}=\pi/8$ (shown in Fig. \ref{geometryupdown}). This parameter set was chosen based on prior work \cite{ball2018} showing that such a tilt angle drives significant intrinsic momentum flux. The electron response is assumed to be adiabatic and we only consider Ion Temperature Gradient (ITG)-type instability and turbulence.
   %The blue curves in Fig. \ref{geometryupdown} gives the flux surface shape and geometric factors for these physical parameters. The corresponding surface shape and geometric factors for elongation geometry ($\kappa=1.5$, $\rchi_{\kappa}=0$, other parameters keeping the same) and circular geometry ($\kappa=1.0$, $\rchi_{\kappa}=0$) are also shown as a comparison.
    
\begin{table}\centering
\caption{\label{updownphysicspara} The physical parameters used in GENE simulations of up-down asymmetric geometries. The electrons are assumed to be adiabatic with $T_e=T_i$.}
% GENE grid parameters for up-down asymmetric simulations. For $\hat{s}=0.1\sim0.8$, inverse aspect ratio $\epsilon=0.18$, $q=1.05\sim4.05$, elongation $\kappa=1.5$, tilt angle $\rchi_k=\pi/8$, $R_0/L_T=6.96$, $R_0/L_n=2.22$.
\footnotesize
\begin{tabular}{@{}ll}
\br\centering
% Simulation Type & $(n_{k_x},n_{k_y},n_z,n_{v_{||}},n_{\mu})$& &$M$\\
Parameter&Value\\
\mr\centering
Magnetic shear $\hat{s}$&$0.1,0.4,0.8$\\
Safety factor $q$&$1.05,2.05,3.05,4.05$\\
Inverse aspect ratio $\epsilon$&$0.18$\\
Elongation $\kappa$&$1.5$\\
Elongation tilt angle $\theta_{\kappa}$&$\pi/8$\\
Temperature gradient $R_0/L_T$&$6.96$\\
Density gradient $R_0/L_n$&$2.22$\\
\br\centering
\end{tabular}\\
\end{table}

\normalsize

\begin{figure}
\centering
\includegraphics[width=0.96\textwidth]{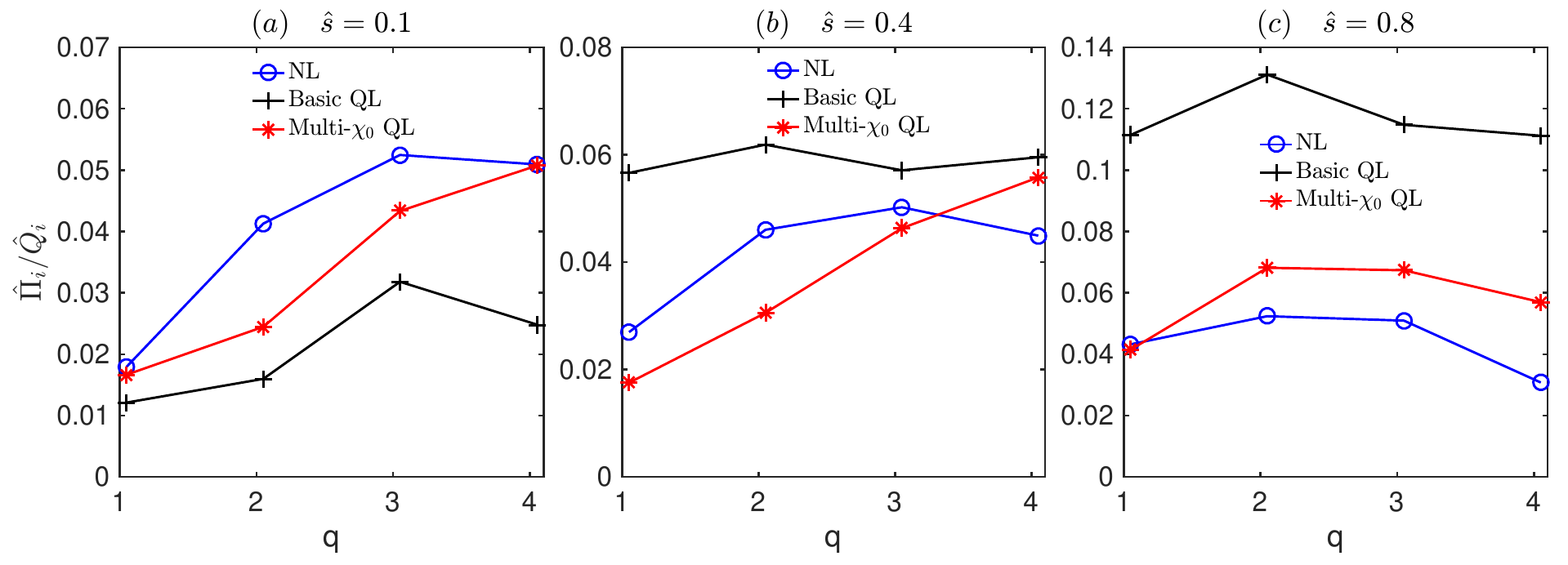}
\caption{\label{Prupdownasy} A comparison of $\hat{\Pi}_i/\hat{Q}_i$ for NL simulations (blue circles), the basic QL estimate (black crosses) and our new multi-$\rchi_0$ QL estimate (red stars) for (a) $\hat{s}=0.1$, (b) $\hat{s}=0.4$ and (c) $\hat{s}=0.8$.}
\end{figure}

Figure \ref{Prupdownasy} shows the results of our benchmark. We compare the flux ratio $\hat{\Pi}_i/\hat{Q}_i$ of both the basic and multi-$\rchi_0$ QL estimates as well as the NL simulations. Here $\hat{\Pi}_i$ and $\hat{Q}_i$ are the normalized toroidal angular momentum flux and heat flux, respectively, where $\hat{\Pi}_i$ is normalized by $c_s^2 m_i n_i R_0(\rho_i/R_0)^2$ and $\hat{Q}_i$ is normalized by $c_s n_i T_i(\rho_i/R_0)^2$, $T_i$ is the ion temperature, $m_i$ is the ion mass, $n_i$ is the ion density, $R_0$ is the major radius, $\rho_i$ is the ion gyroradius, $c_s=\sqrt{T_e/m_i}$ is the sound speed, and $T_e$ is the electron temperature. Figure \ref{Prupdownasy} shows that the basic QL model which considers only the ballooning angle $\rchi_0=0$, does not match well with NL simulations (compare black and red lines), while the multi-$\rchi_0$ model agrees significantly better with NL simulations (compare blue and red lines). The average deviation of the multi-$\rchi_0$ QL model from NL simulations is about $20\%$, while the basic QL model has an average deviation of more than $100\%$. This demonstrates the reliability of our new multi-$\rchi_0$ QL model. 

\begin{figure}
\centering
\includegraphics[width=0.92\textwidth]{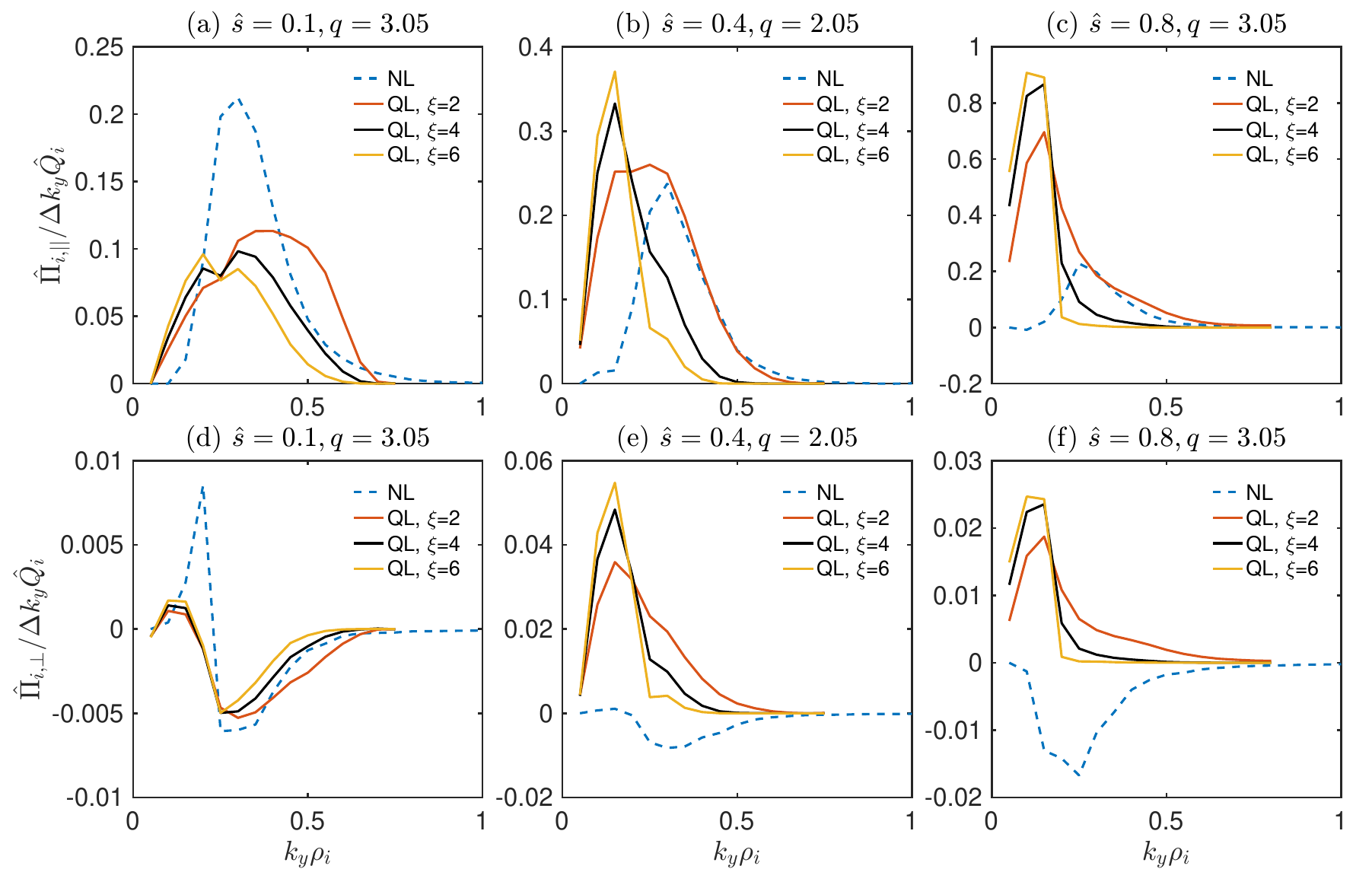}
\caption{\label{spectrumnexc=1} Spectra of the parallel $\hat{\Pi}_{i,||}$ (top row) and perpendicular $\hat{\Pi}_{i,\perp}$ (bottom row) momentum fluxes for NL simulations (blue dashed) and basic QL estimates with $\xi=2$ (red), $\xi=4$ (black) and $\xi=6$ (yellow) for representative up-down asymmetric cases without external flow shear. Here $\Delta k_y\rho_i=0.05$.}
\end{figure}

In order to gain more confidence, we also compare the $k_y$ spectrum of the parallel component of the toroidal angular momentum flux $\hat{\Pi}_{i,||}$ as well as of the perpendicular component $\hat{\Pi}_{i,\perp}$ between NL simulations and the QL estimates. Figure \ref{spectrumnexc=1} shows the comparison between spectra from the basic QL model and the reference NL simulations. As with the total fluxes, they do not agree well. This is because the basic QL model only considers the single $\rchi_0=0$, which does not capture all the important ballooning modes in NL simulations when the geometry is up-down asymmetric. 
%fails to consider the ballooning mode that is most important in NL simulations. 
%(a)-(c) compares the $\Pi_{i,||}$ spectrum between NL simulations and QL estimates with $M=1$ for $\xi=2,4,6$. (d)-(f) compares the corresponding $\Pi_{i,\perp}$ spectrum.

\begin{figure}
\centering
\includegraphics[width=0.92\textwidth]{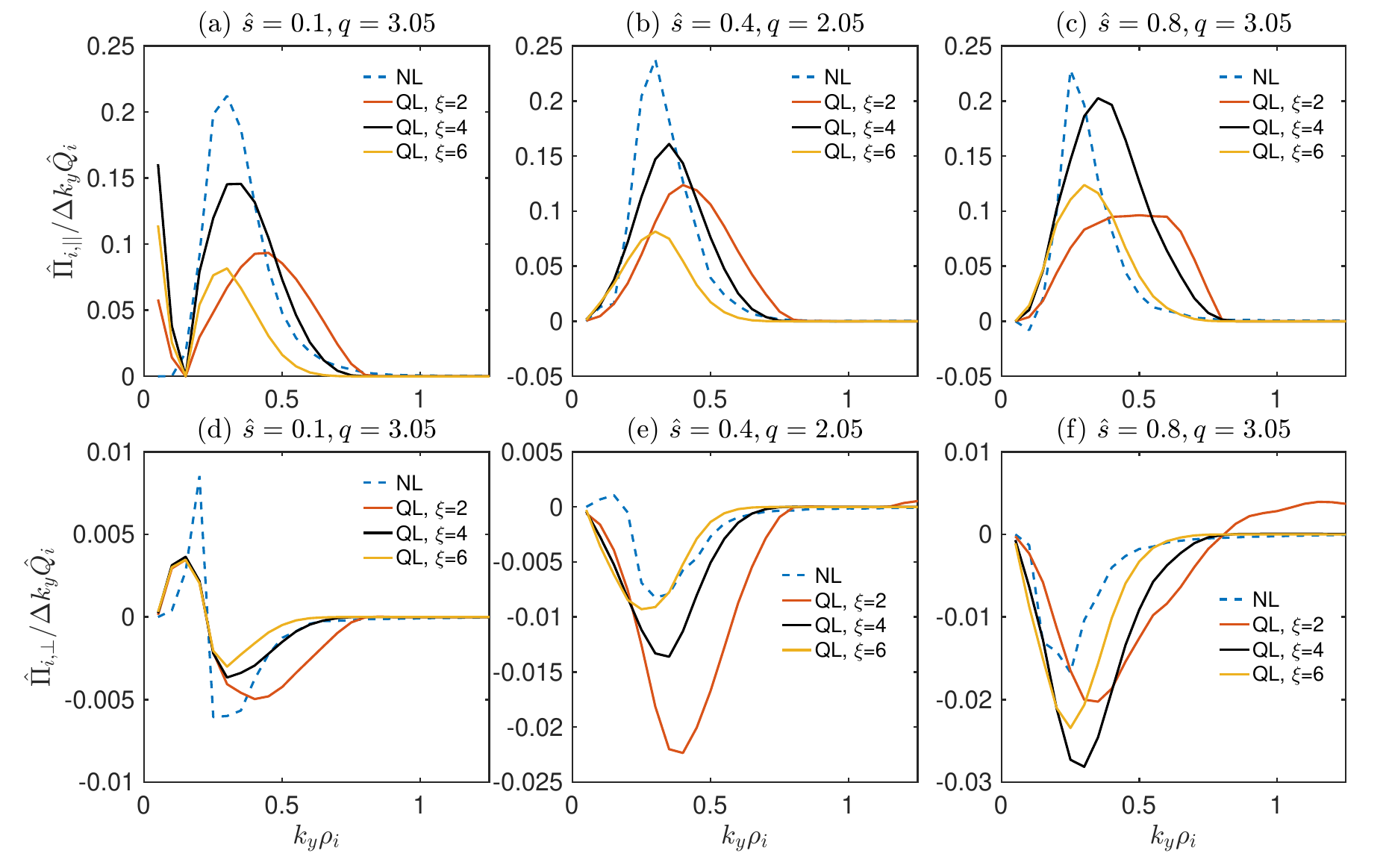}
\caption{\label{spectrumnexc=8} Same as Fig. \ref{spectrumnexc=1}, but using the multi-$\rchi_0$ QL model.}
\end{figure}

The evidence for this is given in Fig. \ref{spectrumnexc=8}, which shows the same comparison between NL simulations and the multi-$\rchi_0$ QL estimate. As we can see, almost all the cases show much improved matches. The agreement is not perfect, but QL models are ultimately expected to provide only estimates. We can also see that $|\hat{\Pi}_{i,||}|\gg|\hat{\Pi}_{i,\perp}|$, so we mainly focus on the best match of $\hat{\Pi}_{i,||}$ estimate to identify the optimal QL model. From these results, the QL estimates obtained for $\xi=4$ are the ones that match best with the NL simulations, which motivates us to use $\xi=4$ in the further development of our model. Note that $\hat{\Pi}_{i,||}$ is positive, which represents diffusive momentum transport, while $\hat{\Pi}_{i,\perp}$ is negative reflecting that it is anti-diffusive \cite{ball2016a}.

%% . The linear simulation with $M=8$ is performed by simply changing the $M$ in the linear runs (without $\rchi_0$ scan) shown in Tab. \ref{updownpara} from $1$ to $8$ (see \ref{AppendixA}). This will include
%The ballooning structure in the figure is calculated by $\langle G_{b,k_y}(\rchi_0,z_b,t)/\text{max}_{\rchi_0,z_b}(G_{b,k_y}(\rchi_0,z_b,t))\rangle_t$, where $G$ could be $\Pi_{i,||}$, $\Pi_{i,\perp}$, $Q_i$ or $\phi$, "b" refers to ballooning space.

\begin{figure}
\centering
\includegraphics[width=0.92\textwidth]{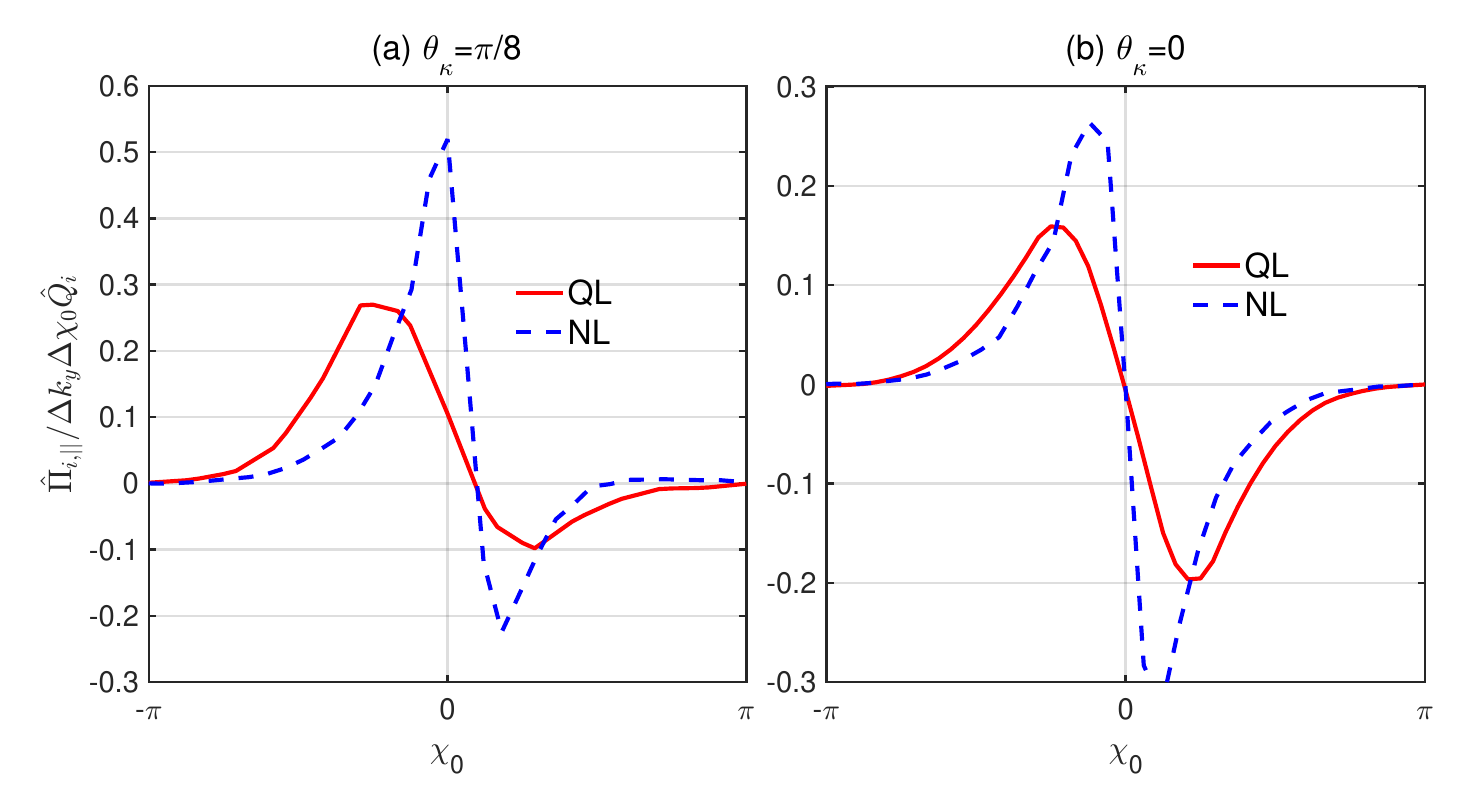}
\caption{\label{ballanglecheck} A comparison of the dependence on ballooning angle $\rchi_0$ for the flux $\hat{\Pi}_{i,||}$ between NL simulations (blue dashed) and the multi-$\rchi_0$ QL model (red solid) for the case with $\epsilon=0.18, R_0/L_T=6.96, q=3.05, \hat{s}=0.8$ and $\kappa=1.5$ for (a) an up-down asymmetric geometry with $\theta_{\kappa}=\pi/8$ and (b) an up-down symmetric geometry with $\theta_{\kappa}=0$.}
\end{figure}
 
\begin{figure}
\centering
\includegraphics[width=0.92\textwidth]{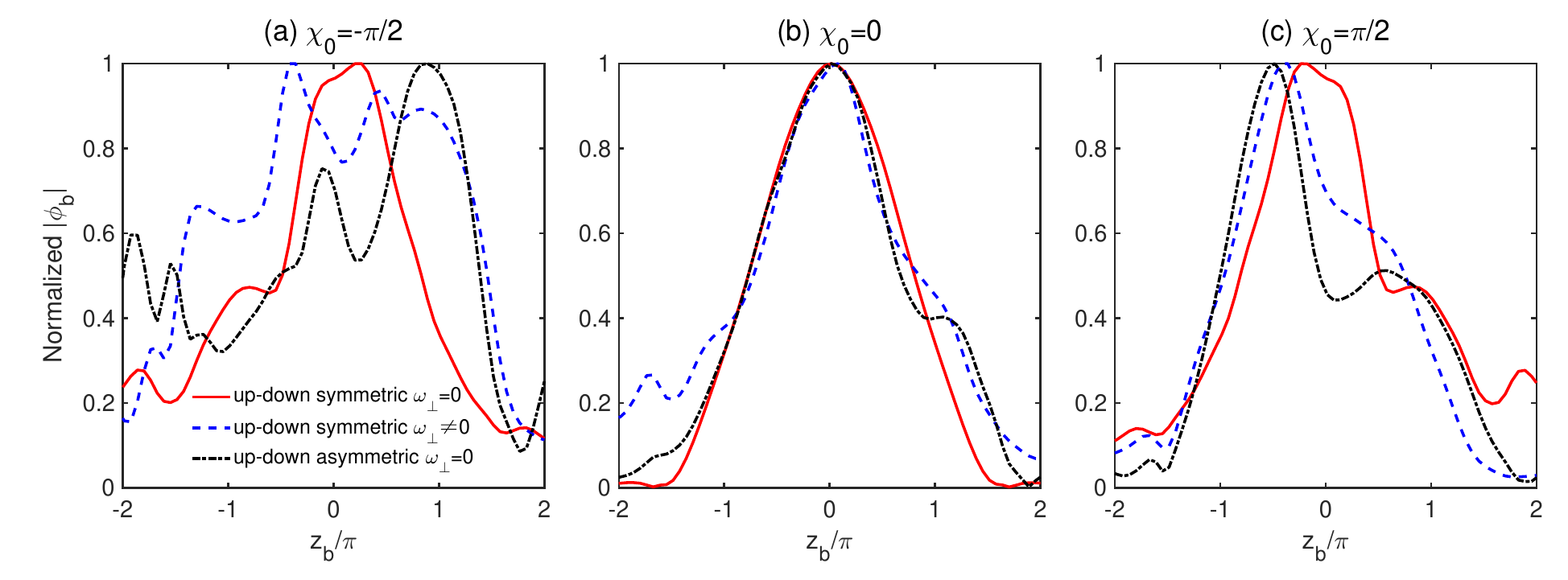}
\caption{\label{linearball3cases} A comparison of the linear ballooning structure with $\epsilon=0.18, R_0/L_T=6.96, q=3.05, \hat{s}=0.8$ \text{and} $\kappa=1.5$ for the up-down symmetric geometry without flow shear (red solid), the up-down symmetric geometry with a flow shear of $\omega_{\perp}R_0/c_s=0.12$ (blue dashed) and the up-down asymmetric geometry with $\theta_{\kappa}=\pi/8$ without flow shear (black dash dotted). Plots are shown for three different ballooning angles (a) $\rchi_0=-\pi/2$, (b) $\rchi_0=0$ and (c) $\rchi_0=\pi/2$.}
\end{figure}
To further demonstrate why using multiple ballooning angles is essential, Fig. \ref{ballanglecheck} shows the ballooning angle dependence of $\hat{\Pi}_{i,||}$ at $k_y\rho_i=0.3$ for the case shown in Fig. \ref{spectrumnexc=8} (c) and a reference case without up-down asymmetry ($\theta_{\kappa}=0$). It is clear from the figure that the toroidal angular momentum flux varies significantly for different ballooning angles. The part with positive ballooning angle contributes negatively, while the part with negative ballooning angle contributes positively to the toroidal angular momentem flux. In the up-down asymmetric case, the positive part is larger than the negative part, resulting in a net positive momentum flux when summing over the ballooning angles, as shown in Figs. \ref{spectrumnexc=1} and \ref{spectrumnexc=8}. In the up-down symmetric case, the positive part cancels with the negative part, giving zero momentum flux (note for the blue dashed line in (b), the positive part does not cancel exactly with the negative part due to unavoidable statistical error in NL simulations). Figure \ref{ballanglecheck} indicates that one has to consider multiple $\rchi_0$ values in order to correctly resolve the ballooning angle dependence of the momentum flux. Figure \ref{linearball3cases} shows a comparison of linear ballooning structures for three different cases: an up-down symmetric geometry without flow shear, an up-down symmetric geometry with flow shear, and an up-down asymmetric geometry without flow shear. The definition of flow shear will be given in Sec. \ref{section3}. As we can see, without any flow shear and no up-down asymmetry, the normalized ballooning structures of the electrostatic field $|\phi_b|$ verify the symmetry $|\phi_b(-\rchi_0,k_y,-z_b)|=|\phi_b(\rchi_0,k_y,z_b)|$, in agreement with Ref. \cite{Camenen_2011NFReview}. In particular, the structure is even with respect to $z_b$ when $\rchi_0=0$. In this case, choosing $\rchi_0=0$ gives reasonable results because it is a good representation of the average ballooning structure of the other ballooning angles. However, with up-down asymmetry or flow shear, the above-mentioned symmetry is broken, so they can only be appropriately described in the QL model by accounting for the contributions from the ballooning structures of multiple $\rchi_0$. 
%the ballooning structure becomes much more complicated and strongly asymmetric. The ballooning structure of ballooning angle $\rchi_0=0$ can no longer represent every ballooning angle in the system, which necessitates the consideration of multiple ballooning angles.

\begin{figure}
\centering
\includegraphics[width=0.96\textwidth]{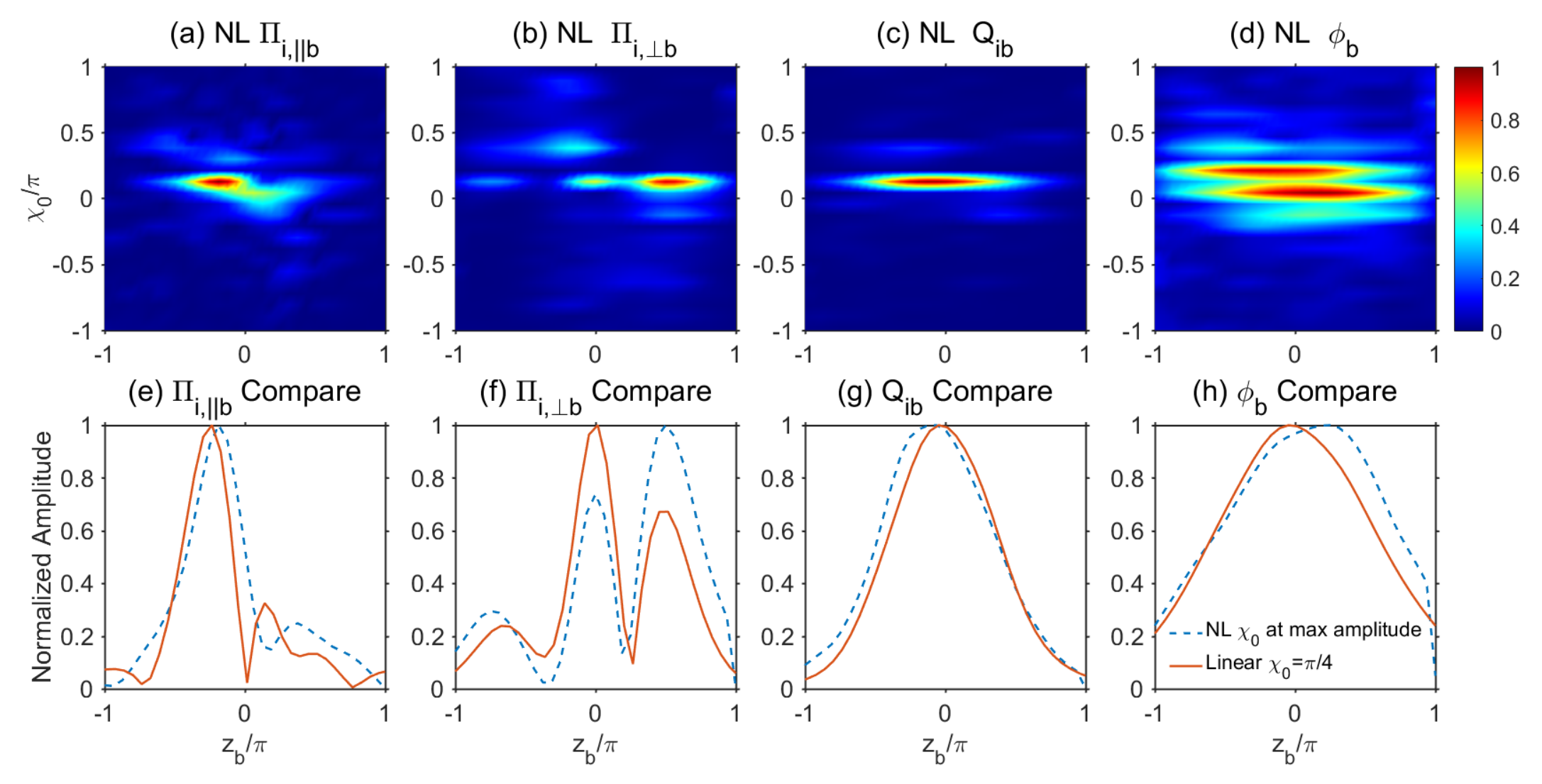}
\caption{\label{ballupdown} Ballooning space structures for the up-down asymmetric geometry with $\hat{s}=0.8$ and $q=3.05$. Sub-plots (a)-(d) give the ballooning space structure for the electrostatic potential $\phi_b$, parallel $\Pi_{i,||b}$ and perpendicular $\Pi_{i,\perp b}$ components of the toroidal angular momentum flux and the the heat flux $Q_{ib}$ from NL simulations. Sub-plots (e)-(h) give comparisons between linear ballooning structures at $\rchi_0=\pi/4$ (which is the fastest growing ballooning mode linearly) and the corresponding NL ballooning structures at $\rchi_0=\pi/8$ with the maximum saturation amplitude. All the plots take $k_y\rho_i=0.15$ mode, which is the dominant mode in the saturated state of the NL simulations.}
\end{figure}

Figure \ref{ballupdown} shows the ballooning structures of the electrostatic field $\phi_b$ and the fluxes $\Pi_{i,||b}$, $\Pi_{i,\perp b}$, $Q_{ib}$ as a function of $\rchi_0$ and $z_b$ in NL simulations for the $\hat{s}=0.8, q=3.05$ case. It also shows a comparison with the corresponding linear simulations using $M=8$, i.e., eight ballooning angles, equally spaced within the range $\rchi_0\in(-\pi,\pi]$. The NL ballooning structures are calculated using a time average over the saturated state and each data set for each considered physical quantity is normalized to its maximum value. As we can see, the peak location of the ballooning structures in NL simulations is not located at $\rchi_0=0$. This is consistent with the fact that the fastest growing mode in the corresponding linear simulations has $\rchi_0=\pi/4$ instead of $\rchi_0=0$. Thus, if a QL model only considers ballooning modes centered at $\rchi_0=0$, it will not capture the most important modes driving the turbulence and associated fluxes. This explains why the basic QL model struggles for up-down asymmetric and low $\hat{s}$ cases. 
%Note that in the NL simulations, the ballooning structure presents the maximum amplitude for a value of the ballooning angle different from $\rchi_0=0$, implying the requirement for more than one ballooning angle in QL estimates.

\section{Extending the QL model to include flow shear}\label{section3}
%So the linear behavior of each $\rchi_0^*$ is $2\pi$ periodic in $\rchi_0$.
In the previous section, we have shown the importance of considering multiple ballooning angles when modelling momentum transport in simulations with up-down asymmetric geometry and low magnetic shear $\hat{s}$. In this section, we further consider cases with background flow shear. The direct effect of perpendicular flow shear on linear eigenmodes is to push their ballooning angle $\rchi_0$ in time according to the relation  
\be\label{eq_floquet}
\rchi_0=\rchi_0^*+\omega_{\perp}(t-t_0)/\hat{s}, 
\ee
%In Eq. \ref{eq_floquet}, choosing constant values for $\rchi_0^*$ and then $k_y$ identifies a linear eigenmode and determines how its ballooning angle $\rchi_0$ value evolves with time. This means that
where $\rchi^*_0$ is the initial ballooning angle at some reference time $t_0$, $\omega_{\perp}=-(r_0/q) \partial\Omega_{tor}/\partial x$ is the $E\times B$ shearing rate consistent with purely toroidal rotation with angular velocity shearing $\partial \Omega_{tor}/\partial t$, $r_0$ is the radial location of the flux tube center, $\Omega_{tor}$ is toroidal angular frequency, and $t$ is time. For all the simulations in this paper with flow shear, we consider purely toroidal flow, resulting from the appropriate combination of parallel and perpendicular ($E\times B$) flow. For convenience, the strength of the flow shear will be quantified by $\omega_{\perp}$. According to Eq. \ref{eq_floquet}, flow shear causes modes to twist with time as long as $\hat{s}\ne0$ \cite{casson2009,NewtonFlowShearUnderstanding2010,ChristenFlowShear2018,ball2019}. Based on Eq. \ref{eq_Coordinate1}, the $k_x$ Fourier modes are linearly coupled such that $\rchi_0$ and $\rchi_0+p2\pi$ (where $p$ is an integer) are part of the same linear eigenmodes (as is in the case for $\omega_{\perp}=0$). With flow shear, these Fourier modes are pushed by the flow shear (based on Eq. \ref{eq_floquet}) and are all covered by a single eigenmode. The time it takes for a Fourier mode to be pushed by the $E\times B$ shear flow to its neighboring linearly coupled Fourier mode is referred to as the Floquet period \cite{NewtonFlowShearUnderstanding2010,ChristenFlowShear2018}. From Eq. \ref{eq_floquet}, we see that the Floquet period is given by $t_F=2\pi\hat{s}/\omega_{\perp}$. In the GENE convention, based on relation $k_{x0}=-k_y\hat{s}\rchi_0$ and Eq. \ref{eq_floquet}, a positive flow shear $\omega_{\perp}$ will push a mode in the negative $k_x$ direction. The long-time evolution of a single linear mode will therefore not just be exponential, but also present modulation with period $t_F$ of its growth rate and frequency as the mode experiences different dynamics at different values of $\rchi_0$ (see Fig. \ref{traceballill} for an illustration). Because of this additional complexity, one actually has to follow the time evolution of linear modes in order to construct a QL model. In this section, we will show how to extend the multi-$\rchi_0$ QL model for the momentum transport to include background flow shear. In combination with flow shear, we will consider challenging parameter regimes including tight aspect ratio, low and high magnetic shear, kinetic electrons, and up-down asymmetric geometry. 
%Before we move on to construct the model, it is important to understand the physics, i.e., the ballooning space structure at both low and high magnetic shear in the presence of flow shear. 

\subsection{Description of the extended QL model with flow shear}\label{modelreal}
\begin{figure}
\centering
\includegraphics[width=0.92\textwidth]{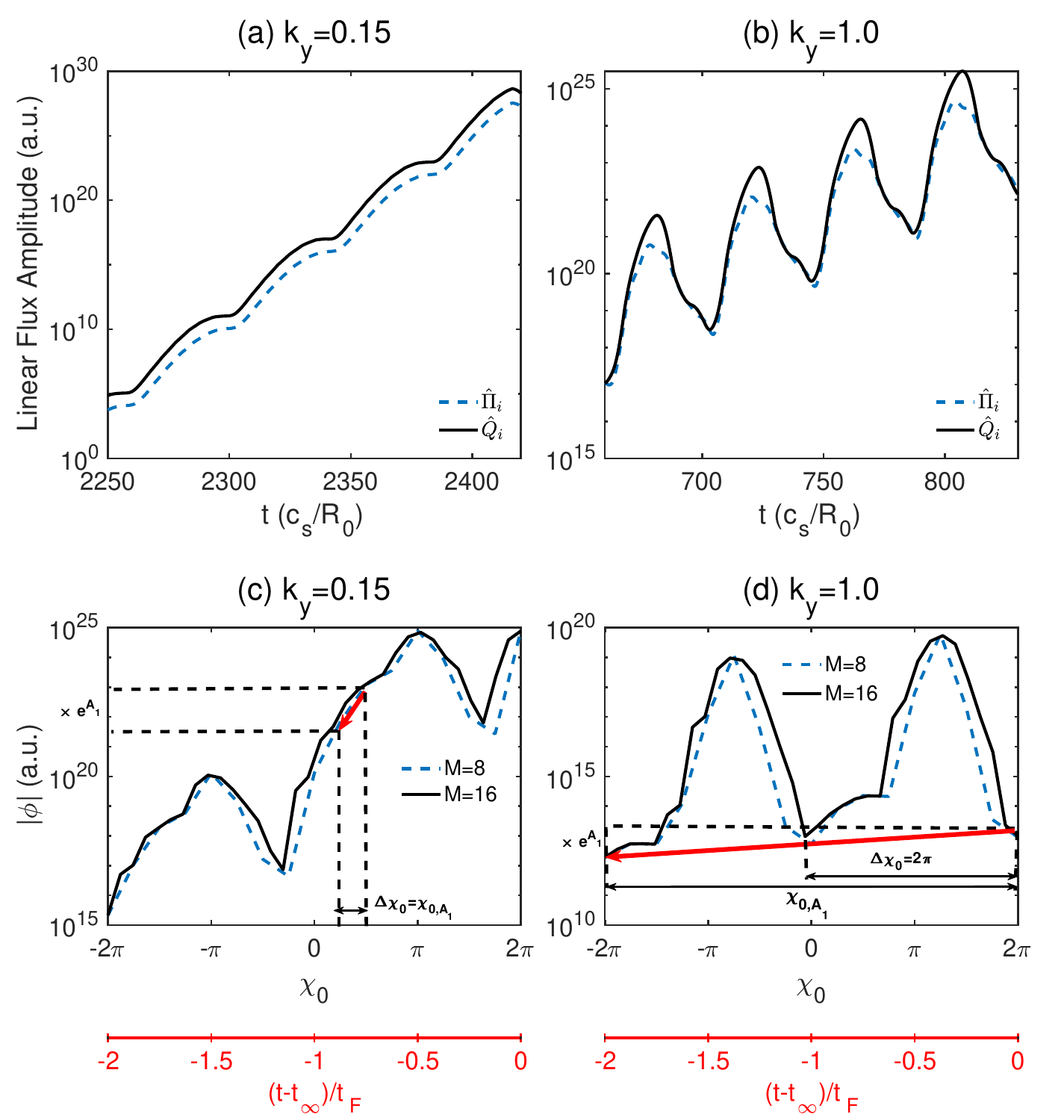}
\caption{\label{traceballill} (a) and (b) show the time evolution of normalized ion toroidal angular momentum flux $\hat{\Pi}_i$ and heat flux $\hat{Q}_i$ for $k_y\rho_i=0.15$ and $k_y\rho_i=1.0$, respectively. (c) and (d) show the time evolution of the electrostatic field amplitude $|\phi|$ for a fast ($k_y\rho_i=0.15$) and a slowly ($k_y\rho_i=1.0$) growing mode, respectively. In the model, $\Delta \rchi_0$ is chosen to be the change in ballooning angle for which the mode amplitude has been reduced by a factor of $e^{A_1}$ (and $A_1=1$ was considered in practice). However, if the amplitude changes slowly, $\Delta\rchi_0$ is limited at a maximum value of $2\pi$, corresponding to a full Floquet period. In (c), $\Delta \rchi_0=\rchi_{0,A_1}$, while in (d), $\Delta \rchi_0=2\pi$. Here $R_0/L_T=10.96, \hat{s}=0.8, q=3.05$ \text{and} $\omega_{\perp}R_0/c_s=0.12$.}
\end{figure}

As mentioned above, the presence of flow shear pushes every ballooning mode along $\rchi_0$. Fortunately, for a given $k_y$, the time evolution of the linear modes (identified by different initial values $\rchi_0^*$ of the ballooning angle) becomes identical given that they experience the same evolution as their ballooning angle $\rchi_0$ gets shifted according to Eq. \ref{eq_floquet}. Thus, to construct a generalization of our previous QL model to include flow shear, it is sufficient to follow a single linear mode throughout its evolution over a Floquet period $t_F$. As the mode passes through each value of $\rchi_0\in (-\pi,\pi]$, we can take that eigenfunction and weight it by an estimate of its amplitude relative to the eigenfunction at other values of $\rchi_0$ and $k_y$. The new generalized QL model that we propose is therefore constructed as follows
\be\label{eq_QL7-0}
F^{QL} = A_0\sum_{\rchi_{0},k_y}w_{fs}^{QL}(\rchi_{0},k_y)F^{L}_{norm}(\rchi_{0},k_y).
\ee
At this level, the relation given by Eq. \ref{eq_QL7-0} for estimating a given flux quantity $F^{QL}$ appears essentially identical to Eq. \ref{eq_QL1-6}, where \say{$fs$} in $w^{QL}_{fs}$ refers to \say{flow shear}. This reflects the fact that we are still just weighting contributions from ballooning mode structures at different values of $k_y$ and $\rchi_0$. The normalized linear flux is given by
\be\label{eq_QL7-1}
F^{L}_{norm}(\rchi_{0},k_y)=\frac{\left\langle F_{b}^{L}\left(\rchi_{0},k_y,z_b,t_{\infty}-\frac{\hat{s}}{\omega_{\perp}}(\rchi^{*}_0-\rchi_{0})\right)\right\rangle_{z_b}}{\text{MAX}_{z_b}\left[\left|\phi_{b}\left(\rchi_{0},k_y,z_b,t_{\infty}-\frac{\hat{s}}{\omega_{\perp}}(\rchi^{*}_{0}-\rchi_{0})\right)\right|^2\right]}, 
\ee
%, using the equation mentioned in Sec. \ref{QLmodelsmall} ($\phi_b(k_{x0},k_y,z_b,t)=\phi(k_x=k_{x0}+2\pi k_y\hat{s}P(z_b),k_y,z=z_b-2\pi P(z_b),t)$)
% \be\label{eq_QL7theta0star}
% \rchi^{*}_{0}=\rchi_0-\omega_{\perp}(t_{\infty}-t)/\hat{s}=-2\pi, 
% \ee
% \be\label{eq_QL7theta0stardiff}
% d\rchi_0=-\omega_{\perp}/\hat{s}dt. 
% \ee
%From Eq. \ref{eq_QL7theta0star} and \ref{eq_QL7theta0stardiff}, we can see that we have chosen the linear mode with a ballooning angle of $\rchi_{0}=-2\pi$ at the end of the simulation $t=t_{\infty}$.
%The $\rchi^{*}_0$ in Eq. \ref{eq_QL7-1} is taken to be a constant $\rchi^{*}_{0}=\rchi_0+\omega_{\perp}(t_{\infty}-t)/\hat{s}=2\pi$ as it simply chooses which of the many identical linear modes to follow ($t_{\infty}$ still denotes the final simulation time). Here we can also see a differential relation $d\rchi_0=\omega_{\perp}/\hat{s}dt$. In the model, for each $k_y$, all the time-dependent physical quantities are calculated for a single ballooning mode characterized by a given $\rchi^{*}_0$ as it moves through different $\rchi_0$ values. In Eq. \ref{eq_floquet}, if we reduce $t$ while holding $\rchi_0^*$ constant, we can trace the ballooning mode $\rchi^{*}_0$ back in time when it had different values of $\rchi_0$. Equivalently, we can change $\rchi_0$ (still holding $\rchi_0^*$ constant) in Eq. \ref{eq_floquet} to find the time at which the linear mode had a given $\rchi_0$. Therefore, 
where $t_{\infty}$ still stands for the final simulation time. In Eq. \ref{eq_QL7-1}, $\rchi_0^*$ is the ballooning angle at $t=t_{\infty}$, while $\rchi_0$, according to Eq. \ref{eq_floquet} with $t_0=t_{\infty}$, is the ballooning angle of the same Floquet mode at time $t=t_{\infty}-(\rchi_0^*-\rchi_0)\hat{s}/\omega_{\perp}$. Note that $F^{L}_{norm}$, as given by Eq. \ref{eq_QL7-1}, is independent of $\rchi_0^*$. For a given Floquet mode $\rchi_0^*$, $\rchi_0$ and $t$ are clearly not independent variables. To ensure close analogy between Eq. \ref{eq_QL1-6} and Eq. \ref{eq_QL7-0}, we will use $\rchi_0$ as the independent variable, but one should remember that summing over $\rchi_0$ in Eq. \ref{eq_QL7-0} is equivalent to integrating over time. Equation \ref{eq_QL7-1} thus provides the normalized flux for the Floquet mode $k_y$ at the phase of its Floquet period where it reached the ballooning angle $\rchi_0$. Therefore, this QL model requires the time evolution of the simulation instead of just looking at the one last time step. %Specifically, we will see that in practice one needs data from the last two Floquet periods (corresponding to $\rchi_0\in(-2\pi,2\pi]$).
%From this we can understand Eq. \ref{eq_QL7-1}. To get the normalized flux for a given value of $\rchi_0$, we take the $\rchi_0^*=2\pi$ linear mode and trace it back in time from the last time step $t_{\infty}$ until it had the desired value of $\rchi_0$.
For a given $k_y$, to weight the contributions from the different $\rchi_0$ values, we use
\be\label{eq_QL7-2}
w_{fs}^{QL}(\rchi_{0},k_y)= 
\left\{
% \begin{matrix}
\begin{aligned}
%+\frac{\omega_{\perp}}{\hat{s}}t,
%\int_{0}^{\Delta\rchi_0(\rchi_0,k_y)} d\rchi_0^{'}\frac{\gamma( \rchi_0+\rchi_0^{'},k_y,t_{\infty}-\frac{\hat{s}}{\omega_{\perp}}(\rchi_0+\rchi_0^{'}-\rchi^*_0))}{\langle k^2_{\perp b}\rangle(\rchi_0+\rchi_0^{'},k_y)}
\Lambda^{\xi} \quad\quad\quad\quad\quad\quad\quad\quad\quad\quad\quad\quad\quad\quad\quad\quad\quad\quad\quad\quad\quad\quad\quad\quad\quad\quad\quad\quad\quad\quad\quad\quad\quad\quad\\       \text{if}\quad \Lambda\equiv\frac{1}{\Delta \rchi_0(\rchi_0,k_y)}\int_{0}^{\Delta\rchi_0(\rchi_0,k_y)} d\rchi_0^{'}\frac{\gamma( \rchi_0-\rchi_0^{'},k_y,t_{\infty}-\frac{\hat{s}}{\omega_{\perp}}(\rchi^*_0-\rchi_{0}+\rchi_0^{'}))}{\langle k^2_{\perp b}\rangle(\rchi_0-\rchi^{'}_0,k_y)}>0\\
0 \quad\text{else},\quad\quad\quad\quad\quad\quad\quad\quad\quad\quad\quad\quad\quad\quad\quad\quad\quad\quad\quad\quad\quad\quad\quad\quad\quad\quad\quad\quad\quad\quad\quad\\
%\text{if} \int_{0}^{\Delta\rchi_0(k_y)} d\rchi_0\frac{\gamma( \rchi_0+d\rchi_0,k_y,t_{\infty}-\frac{\hat{s}}{\omega_{\perp}}(\rchi_{0}+d\rchi_0-\rchi^*_0))}{\langle k^2_{\perp}\rangle(\rchi_0+d\rchi_0,k_y)}<0,\xi=4
\end{aligned}
% \end{matrix}
\right.
\ee
%However, Eq. \ref{eq_QL7-2} actually reduces to Eq. \ref{eq_QL1-3} in the limit of weak flow shear and preserves the same spirit.
where $\Lambda$ is defined in this equation. At first glance, Eq. \ref{eq_QL7-2} looks significantly different from Eq. \ref{eq_QL1-3}, the analogue expression in our QL model without flow shear. The role of the QL weight is to estimate the average amplitude of the mode in NL simulations. Without flow shear, this is done using the metric $\gamma/\langle k_{\perp b}^2\rangle$ (see Eq. \ref{eq_QL1-4}). We want to achieve something similar here for $\omega_{\perp}\ne 0$, but $\gamma$ and $\langle k_{\perp b}^2\rangle$ change with time. One could simply use the instantaneous growth rate, which can be calculated with
\be\label{eq_QL7-3-1}
\gamma\left(\rchi_0,k_y,t\right)=\frac{d}{dt}\ln{\left[\phi_b(\rchi_0,k_y,z_b,t)\right]}.
\ee
In practice, this instantaneous growth rate of the Floquet mode is estimated with finite differences using Eq. \ref{eq_floquet} according to
\be\label{eq_QL7-3}
\gamma\left(\rchi_{0},k_y,t_{\infty}-\frac{\hat{s}}{\omega_{\perp}}(\rchi^*_0-\rchi_{0})\right)=\frac{\omega_{\perp}}{\hat{s}\delta\rchi_0}\ln\left(\frac{|\phi_{b}(\rchi_{0},k_y,z_{b0},t_{\infty}-\frac{\hat{s}}{\omega_{\perp}}(\rchi^*_0-\rchi_{0}))|}{|\phi_{b}(\rchi_{0}-\delta\rchi_0,k_y,z_{b0},t_{\infty}-\frac{\hat{s}}{\omega_{\perp}}(\rchi^*_0-\rchi_{0}+\delta\rchi_0))|}\right),
\ee
where $\delta\rchi_0$ is the spacing between the ballooning angles considered in the simulations. Consistent with Eq. \ref{eq_QL1-4}, the instantaneous average perpendicular wavenumber for each ballooning angle $\rchi_0$ is given by
\be\label{eq_QL7-4}
\langle k_{\perp b}^2\rangle(\rchi_0,k_y)=\frac{\langle k_{\perp b}^2(\rchi_0,k_y,z_b)|\phi_{b}(\rchi_0,k_y,z_b,t_{\infty}-\frac{\hat{s}}{\omega_{\perp}}(\rchi^*_0-\rchi_{0}))|^2\rangle_{z_b}}{\langle |\phi_{b}(\rchi_0,k_y,z_b,t_{\infty}-\frac{\hat{s}}{\omega_{\perp}}(\rchi^*_0-\rchi_{0}))|^2\rangle_{z_b}}.
\ee
%This is important in NL simulations as a mode that has been growing for a long time will still have a large amplitude, even if it just stopped growing.
However, this does not take into account the history of the Floquet mode prior to it reaching a given $\rchi_0$. We thus take an average of $\gamma/\langle k_{\perp b}^2\rangle$ over $\rchi_0$ to incorporate the prior history of the evolving eigenmodes. Similar to considerations in Ref. \cite{Roach_2009}, the averaging window $\Delta \rchi_0$ in Eq. \ref{eq_QL7-2} is determined by the following relations
\be\label{eq_QL7-2-2}
\left\{
\begin{aligned}
\ln{\left(\frac{|\phi_{b}(\rchi_{0},k_y,z_{b0},t_{\infty}-\frac{\hat{s}}{\omega_{\perp}}(\rchi^*_0-\rchi_{0}))|}{|\phi_{b}(\rchi_{0}-\rchi_{0,A_1}(\rchi_0,k_y),k_y,z_{b0},t_{\infty}-\frac{\hat{s}}{\omega_{\perp}}(\rchi^*_{0}-\rchi_{0}+\rchi_{0,A_1}(\rchi_0,k_y)))|}\right)}=A_1\\ 
A_1=O(1)\approx 1\quad\quad\quad\quad\quad\quad\quad\quad\quad\quad\quad\quad\quad\quad\quad\quad\quad\quad\quad\quad\quad\quad\quad\quad\quad \\
\Delta \rchi_0(\rchi_0,k_y)=\text{MIN}(\rchi_{0,A_1}(\rchi_0,k_y),2\pi),\quad\quad\quad\quad\quad\quad\quad\quad\quad\quad\quad\quad\quad\quad\quad\quad\\
\end{aligned}
\right.
\ee
where $z_{b0}=0$ is the center of ballooning space, $\rchi_{0,A_1}$ is the backward shift in the ballooning angle $\rchi_0$ at which the mode amplitude was lower by a factor of $e^{A_1}$ and both $\rchi_{0,A_1}$ and $\Delta \rchi_0$ are calculated from this equation. If the mode was not strongly growing and one has to go back more than a Floquet period for a decay by such a factor, one takes $\Delta\rchi_0=2\pi$, which corresponds to a Floquet period $t_F=2\pi\hat{s}/\omega_{\perp}$. Otherwise, one takes $\Delta \rchi_0=\rchi_{0,A_1}$. \textcolor{black}{In practice, we choose $A_1=1$ as it is actually a good measure of the variation of mode fluctuation amplitudes in our NL simulations. The mode fluctuation amplitude is defined by $A(\rchi_0,k_y)=\text{ln}[(\langle\phi_b\rangle_t+\text{SD}_t(\phi_b))/(\langle\phi_b\rangle_t-\text{SD}_t(\phi_b))]$, where $\text{SD}_t(\phi_b)$ is the standard deviation of the time variation of $\phi_b$. We checked that for different $\rchi_0$ and $k_y$ in different NL simulations, $A$ is an $O(1)$ quantity. Note that if the mode is growing more than one e-fold within a Floquet period ($\gamma t_F=\gamma 2\pi\hat{s}/\omega_{\perp}>1$), i.e., if it is either a fast growing mode and/or $\omega_{\perp}/\hat{s}$ is small, the average in Eq. \ref{eq_QL7-2} is over a small ballooning angle interval $\Delta\rchi_0$ and tends to reduce to the instantaneous growth of this mode. Consequently, for $\omega_{\perp}\to 0$, the flow shear QL model reduces to the multi-$\rchi_0$ model.} On the other hand, even if the instantaneous growth rate of a ballooning mode is very small or negative, as long as the average growth rate (weighted by $1/\langle k_{\perp b}^2\rangle$) is positive over the last Floquet period, we still take it into account. This approach is illustrated in Fig. \ref{traceballill}, where we show the time evolution of the linear fluxes and the ballooning eigenmode for a typical simulation case. In subfigure (c), the time trace of the amplitude of a fast-growing ballooning mode is shown, so $\Delta\rchi_0$ is taken to be $\rchi_{0,A_1}$ in Eq. \ref{eq_QL7-2-2} because the mode grows more than one e-fold within one Floquet period. In subfigure (d), on the other hand, the mode grows less than one e-folding over a Floquet period and $\Delta\rchi_0$ is taken to be $2\pi$. The time axis that maps to the $\rchi_0$ axis according to Eq. \ref{eq_floquet} is also shown. The convergence of the ballooning mode evolution is also verified by increasing the number $M$ of considered ballooning angles $\rchi_0$ in the interval $(-\pi,\pi]$ from $8$ to $16$ in the linear GENE simulations. Finally, the average over the ballooning space for estimating both $F^L_{norm}$ and $\langle k^2_{\perp b}\rangle$ according to Eqs. \ref{eq_QL7-1} and \ref{eq_QL7-4} is in fact taken only over $[-3\pi,3\pi)$
\be\label{eq_QL7-5}
\langle...\rangle_{z_b}=\frac{\int^{3\pi}_{-3\pi}dz_b(...)J_b(z_b)}{\int^{3\pi}_{-3\pi}dz_b J_b(z_b)}.
\ee
This is different from the previous models, but will be explained in the next section. Equations \ref{eq_QL7-0} to \ref{eq_QL7-5} constitute what we will call the \say{flow shear} QL model. This model is more computationally expensive than the \say{basic} and \say{multi-$\rchi_0$} QL models introduced in Sec. \ref{section2} because it requires a frequent data output from GENE simulations, especially when $\hat{s}/\omega_{\perp}$ is small and $t_F$ is short. This is because a fixed number of snapshots, corresponding to the state of the eigenmode as it reaches each of the $M$ considered ballooning angles, is required within each Floquet period. In order to further improve computational efficiency, we developed a way to obtain the required data for the flow shear QL model from a single snapshot. This method is presented in \ref{AppendixB}. Importantly, the flow shear QL model can be proven to be reduced to the multi-$\rchi_0$ model if we consider the limit $\omega_{\perp}\to 0$ and sum over all the ballooning angles $\rchi_0\in (-\pi,\pi]$, which in turn reduces to the basic QL model if we include only the $\rchi_0$ ballooning mode. Now we will move on to explain the physical reason for limiting ballooning space in Eq. \ref{eq_QL7-5}.

% The expression for the instantaneous growth rate, $\gamma$, is
% \be\label{eq_QL7-3}
% \gamma\left(\rchi_{0},k_y,t_{\infty}-\frac{\hat{s}}{\omega_{\perp}}(\rchi_{0}-\rchi^*_0)\right)=\gamma\left(\rchi_{0},k_y\right)=\frac{1}{d\rchi_0}\ln\left(\frac{|\phi_{b}(\rchi_{0},k_y,z_{b0},t_{\infty}-\frac{\hat{s}}{\omega_{\perp}}(\rchi_{0}-\rchi^*_0))|}{|\phi_{b}(\rchi_{0}+d\rchi_0,k_y,z_{b0},t_{\infty}-\frac{\hat{s}}{\omega_{\perp}}(\rchi_{0}+d\rchi_0-\rchi^*_0))|}\right).
% \ee
% The expression for $\langle k^2_{\perp b}\rangle(\rchi_0,k_y)$ is given by
% \be\label{eq_QL7-4}
% \langle k_{\perp b}^2\rangle(\rchi_0,k_y)=\frac{\langle k_{\perp b}^2(\rchi_0,k_y,z_b)|\phi_{b}(\rchi_0,k_y,z_b,t_{\infty}-\frac{\hat{s}}{\omega_{\perp}}(\rchi_{0}-\rchi^*_0))|^2\rangle_{z_b}}{\langle |\phi_{b}(\rchi_0,k_y,z_b,t_{\infty}-\frac{\hat{s}}{\omega_{\perp}}(\rchi_{0}-\rchi^*_0))|^2\rangle_{z_b}},
% \ee
% where $k_{\perp b}^2(\rchi_0,k_y,z_b)$ is calculated by transferring $k_{\perp}^2(k_x,k_y,z)$ to the ballooning space. The average over the ballooning space is taken to be only from $-3\pi$ to $3\pi$

% \be\label{eq_QL7-5}
% \langle...\rangle_{z_b}=\frac{\int^{3\pi}_{-3\pi}dz_b(...)J(z_b)}{\int^{3\pi}_{-3\pi}dz_b J(z_b)}.
% \ee

\subsection{Comparing QL and NL ballooning space structure for $\omega_{\perp}\ne 0$}\label{ballspacecomparesec}
\begin{figure}
\centering
\includegraphics[width=0.92\textwidth]{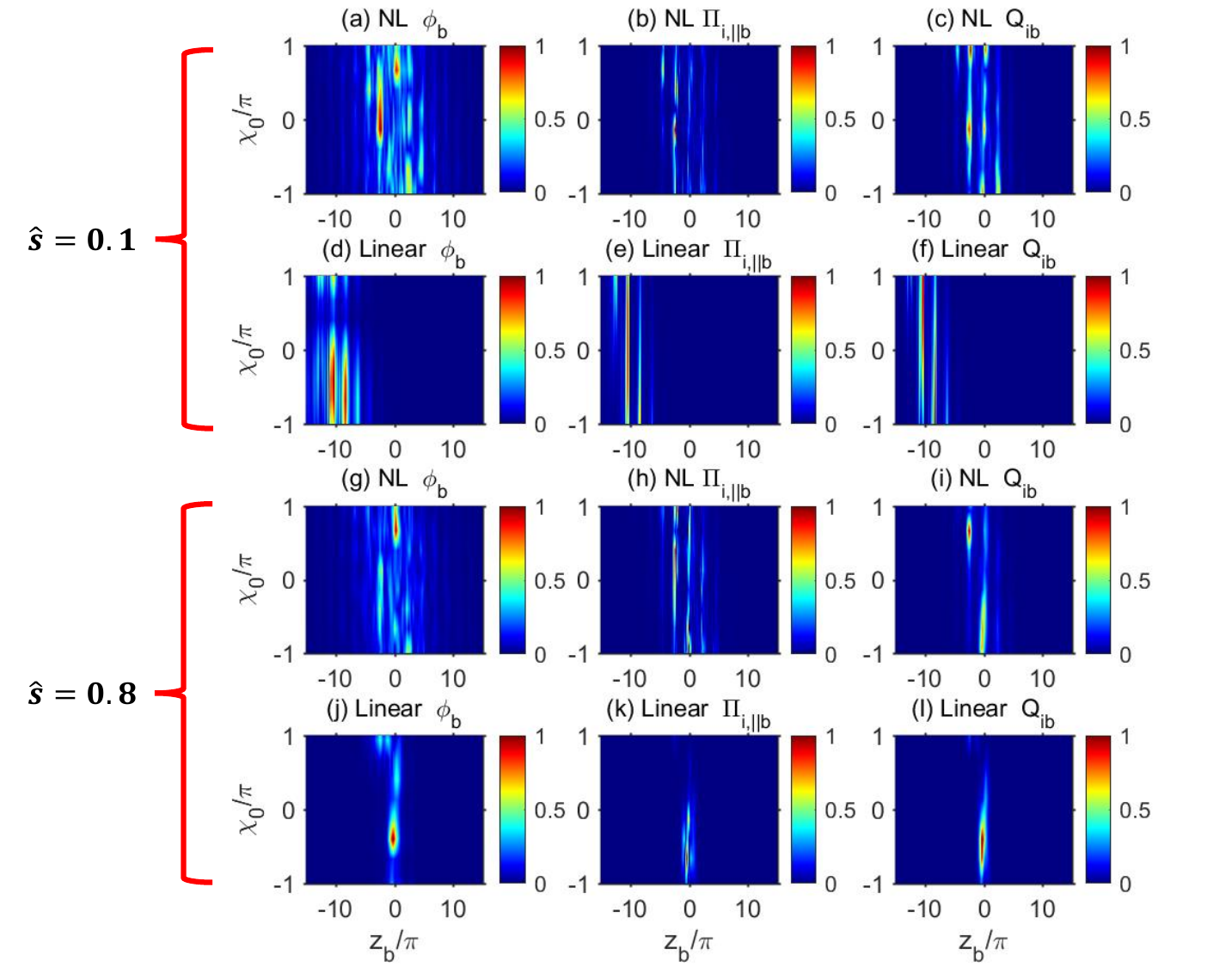}
\caption{\label{balltightshat0.10.8} Ballooning space structures for tight aspect ratio $\epsilon=0.36$, $R_0/L_T=10.96$, $q=3.05$ with flow shear $\omega_{\perp}R_0/c_s=0.12$ and magnetic shear $\hat{s}=0.1$ (for NL (a)-(c) and linear (d)-(f) simulations) as well as $\hat{s}=0.8$ (for NL (g)-(i) and linear (j)-(l) simulations). The first column shows the amplitude of electrostatic field $\phi_b$. The second column shows the parallel component of toroidal angular momentum flux $\Pi_{i,||b}$. The third column shows the heat flux $Q_{ib}$. All the plots are for  $k_y\rho_i=0.15$ mode, which is the dominant one in the NL $|\phi|$ spectrum. The physical parameters are shown in Tab. \ref{tightphysicspara}.}
\end{figure}

In this subsection, we will compare the ballooning structure of fluctuations in linear and NL simulations with non-zero flow shear $\omega_{\perp}\ne 0$ for cases with low and with high magnetic shear $\hat{s}$. This demonstrates the physical reason for limiting ballooning space to be $z_b\in [-3\pi,3\pi)$ in Eq. \ref{eq_QL7-5}. We thus consider the two representative cases, $\hat{s}=0.1$ and $\hat{s}=0.8$, which are shown in Fig. \ref{balltightshat0.10.8}. As we can see, for the high magnetic shear case, the flow shear is not able to push the ballooning structure far away from the central outboard midplane ($z_b=0$) in either the linear or NL simulations. The peak in the mode structure always stays centered around $z_b=0$. However, in linear simulations at low magnetic shear, the location of the maximum of the ballooning structure has been pushed approximately five poloidal turns ($z_b\approx -10\pi$) away from the central location of ballooning space, towards negative values since $\omega_{\perp}>0$. This does not match the NL simulation, which stays centered around $z_b=0$ as in the high magnetic shear simulations. If one does not correct for this discrepancy between the linear and NL results, it will cause significant disagreement between the QL model and the NL simulations. This is achieved by limiting the ballooning space average to $z_b\in [-3\pi,3\pi)$, \textcolor{black}{which is roughly how far the turbulence shifts in $z_b$ in the NL low magnetic shear cases in Fig. \ref{balltightshat0.10.8}.} In this way, one forces the QL model to focus on the modes that are nonlinearly important, which results in a more accurate estimate (see Sec. \ref{tightadiabaticbenchmark}). \ref{AppendixC} presents a physically motivated estimate of how far the modes are advected in ballooning space as well as the limits of applicability that this creates for our QL model.
%As we will show later, we propose to increase the velocity space resolution and limit the ballooning space to only consider several Poloidal turns and obtained good match between NL simulation results and our QL estimates. 

%Therefore, the physical reason (given in detail in \ref{AppendixC}) of limiting the ballooning space average to be between $-3\pi$ to $3\pi$ is because for cases with low magnetic shear and large flow shear, the linear modes are pushed too far along the field line in ballooning space $z_b$ compared to the NL simulations. Therefore, by limiting the ballooning space, we force the QL model to focus on the modes that are nonlinearly important, which results in a more accurate QL estimate (see Sec. \ref{tightadiabaticbenchmark}). \ref{AppendixC} presents a physically motivated estimate of how far the modes are advected in ballooning space as well as the limits of applicability that this creates for our QL model.

\subsection{QL model benchmarking with nonlinear simulations including flow shear}\label{tightadiabaticbenchmark}
We have benchmarked our newly developed flow shear QL model against NL GENE simulations considering tight aspect ratio, circular geometry, and non-zero flow shear. The physical parameters for the benchmark are summarized in Tab. \ref{tightphysicspara}, where we fix the strength of the flow shear $\omega_{\perp}R_0/c_s=0.12$, the aspect ratio $\epsilon=0.36$, the density gradient $R_0/L_n=2.22$, but scan magnetic shear $\hat{s}$, safety factor $q$, and temperature gradient $R_0/L_T$. Electrons are forced to respond adiabatically. The simulation grid parameters are given in Tab. \ref{tightpara} in \ref{AppendixA}. The numerical grid for NL simulations is the same as for the up-down asymmetric simulations and is given in Tab. \ref{updownpara}. Note that here we have increased the number of considered ballooning angles to $M=8$ in linear simulations based on the discussion of our multi-$\rchi_0$ QL model in Sec. \ref{updownbenchmark}. Additionally, as a result of the box quantization condition \cite{BeerBallooingCoordinates1995} $L_x=M/(\Delta k_y\hat{s})$, the low magnetic shear cases require more radial Fourier modes. Indeed, as one decreases $\hat{s}$, $L_x$ becomes larger, which then necessitates more grid points to maintain the same radial resolution. Additionally, larger maximum values of the velocity space grid along $v_{||}$ and $\mu$ are required for low $\hat{s}$ to ensure that information is able to travel along the field line. This is because, as we decrease magnetic shear, the Floquet period $t_F=2\pi\hat{s}/\omega_{\perp}$ decreases, which means that turbulent structures need to travel along field lines faster to stay at the outboard midplane \cite{NewtonFlowShearUnderstanding2010}. This velocity is estimated by $v_{||}\sim qR/t_F\sim qR\omega_{\perp}/\hat{s}$, which should exist within the simulation grid. This effect is primarily important for linear simulations, as NL dynamics more efficiently transfer information, as reflected by the shorter NL decorrelation time $\tau_{NL}$ \cite{casson2009,NewtonFlowShearUnderstanding2010,ChristenFlowShear2018,ball2019}. 

% &$(192,64,32,32,9)$&$10\hat{s}k_y/k_{y,min}$

\begin{table}\centering
\caption{\label{tightphysicspara} The physical parameters used in GENE simulations with circular geometry, non-zero flow shear $\omega_{\perp}R_0/c_s=0.12$, adiabatic electrons with $T_e=T_i$, tight aspect ratio $\epsilon=0.36$, and $R_0/L_n=2.22$.}
%x_0/q_0\partial|\Omega_{tor}|/\partial x R_0/c_s=
\footnotesize
\begin{tabular}{@{}llll}
\br
Simulation Type &$\hat{s}$&$q$&$R_0/L_T$\\
\mr\centering
Nonlinear&$0.1,0.4,0.8$&$2.05,3.05,4.05$&$10.96$\\
Nonlinear&$0.8$&$2.25,3.25,4.25$&$6.96$\\
Nonlinear&$0.4,0.8$&$2.25,3.25,4.25$&$5.06$\\
\mr\centering
Linear (normal $\hat{s}$)&$0.4, 0.8$&$2.05,3.05,4.05$&$10.96$\\
Linear (normal $\hat{s}$)&$0.8$&$2.25,3.25,4.25$&$6.96$\\
Linear (normal $\hat{s}$)&$0.4, 0.8$&$2.25,3.25,4.25$&$5.06$\\
\mr\centering
Linear (low $\hat{s}$)&$0.1$&$2.05,3.05,4.05$&$10.96$\\
% Linear(low$\hat{s}$)&$(768,1,32,64,18)$&$8$&$0.1$&$0.36$&$2.25,3.25,4.25$&$0.12$&$6.96$\\
% Linear(low$\hat{s}$)&$(768,1,32,64,18)$&$8$&$0.1$&$0.36$&$2.25,3.25,4.25$&$0.12$&$5.06$\\
\br\centering
\end{tabular}\\
% $^{a} ExBrate=x_0/q_0\partial|\Omega_{tor}|/\partial x R_0/c_s$.

\end{table}
\normalsize

\begin{figure}
\centering
\includegraphics[width=0.92\textwidth]{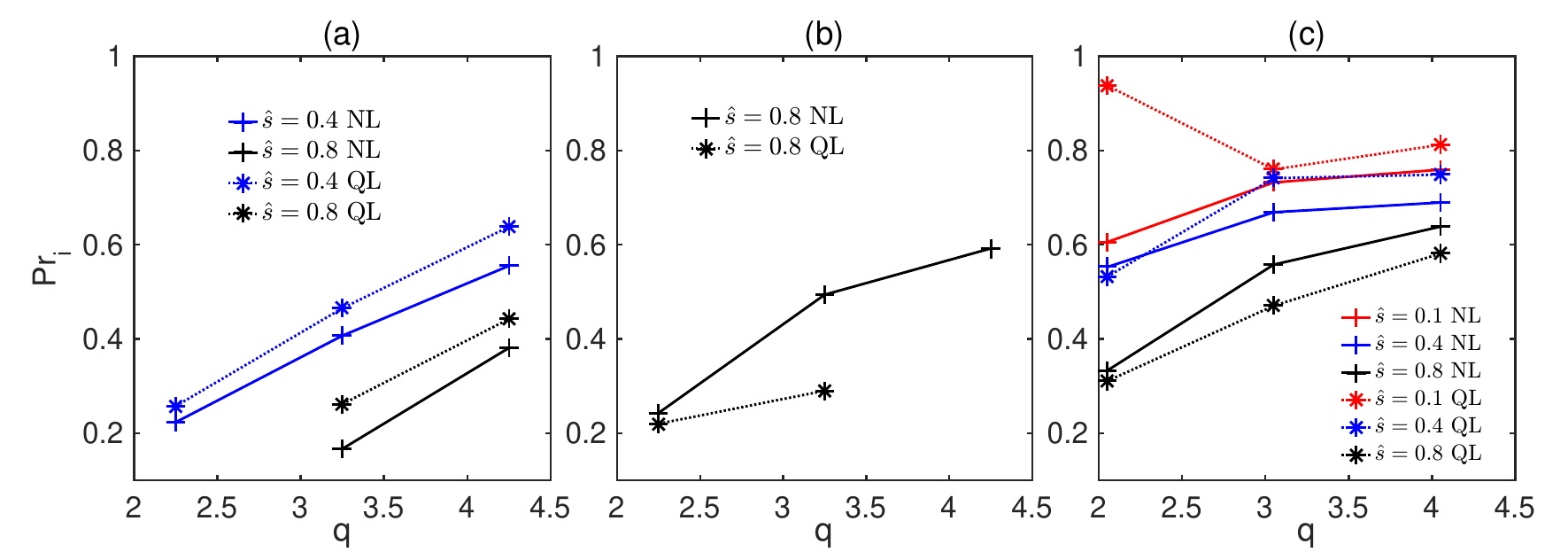}
\caption{\label{tightPrsum} A comparison of the Prandtl number $\text{Pr}_i$ from NL simulations (solid) and the flow shear QL estimate (dashed) for tight aspect ratio $\epsilon=0.36$ cases with temperature gradients of (a) $R_0/L_T=5.06$, (b) $R_0/L_T=6.96$, and (c) $R_0/L_T=10.96$.}
\end{figure}

Figure \ref{tightPrsum} shows a comparison of the Prandtl number obtained with NL simulations and the flow shear QL model for all the cases considered in Tab. \ref{tightphysicspara}. With the standard gyroBohm normalizations considered in GENE, the ion Prandtl number $\text{Pr}_i$ is calculated according to
\be\label{eq_Prandtlnumber}
\text{Pr}_i=\frac{\hat{\Pi}_i}{\hat{Q}_i}\frac{R_0}{L_T}\frac{\epsilon}{q\omega_{\perp}}\frac{c_s}{R_0}.
\ee
The comparison between QL simulations and NL GENE simulations shows a good match in general, except for a few cases that are close to marginal stability (e.g., the case with $R_0/L_T=10.96$, $\hat{s}=0.1$, $q=2.05$ shown in Fig. \ref{tightPrsum} (c)). We see that the Prandtl number increases with $q$, which is consistent with previous work on the low momentum diffusivity regime \cite{ball2016a,Ben_2019}. We also observe that the Prandtl number increases with the temperature gradient $R_0/L_T$ and decreases with magnetic shear $\hat{s}$. One thus concludes that, a low Prandtl number ($\text{Pr}_i\approx0.2$) can be obtained at tight aspect ratio, low safety factor, high magnetic shear, and low temperature gradient. These dependencies are also fully captured in our QL model. The average deviation of the QL model from the actual NL simulations is about $25\%$. This is quite acceptable, since the primary purpose of our QL model is to obtain correct trends for the Prandtl number, such that large parameter scans can be carried out to identify interesting regimes that can be verified by NL studies. 
%The most significant mismatch, the case in Fig. \ref{tightPrsum} (c) is likely due to the turbulence being very close to marginal stability, e.g. the case with $\hat{s}=0.1, q=1.05, R_0/L_T=10.96$ (not shown in the figure) is stable to turbulence. 

%Our QL model, on the other hand, requires a fully developed turbulence in order for the mixing length rule to work. Another possibility is that we didn't fully capture the NL coupling between different $k_x$ modes \cite{Whelan2018QL}. A more detailed analysis to this point will be our future investigations.

\begin{figure}
\centering
\includegraphics[width=0.92\textwidth]{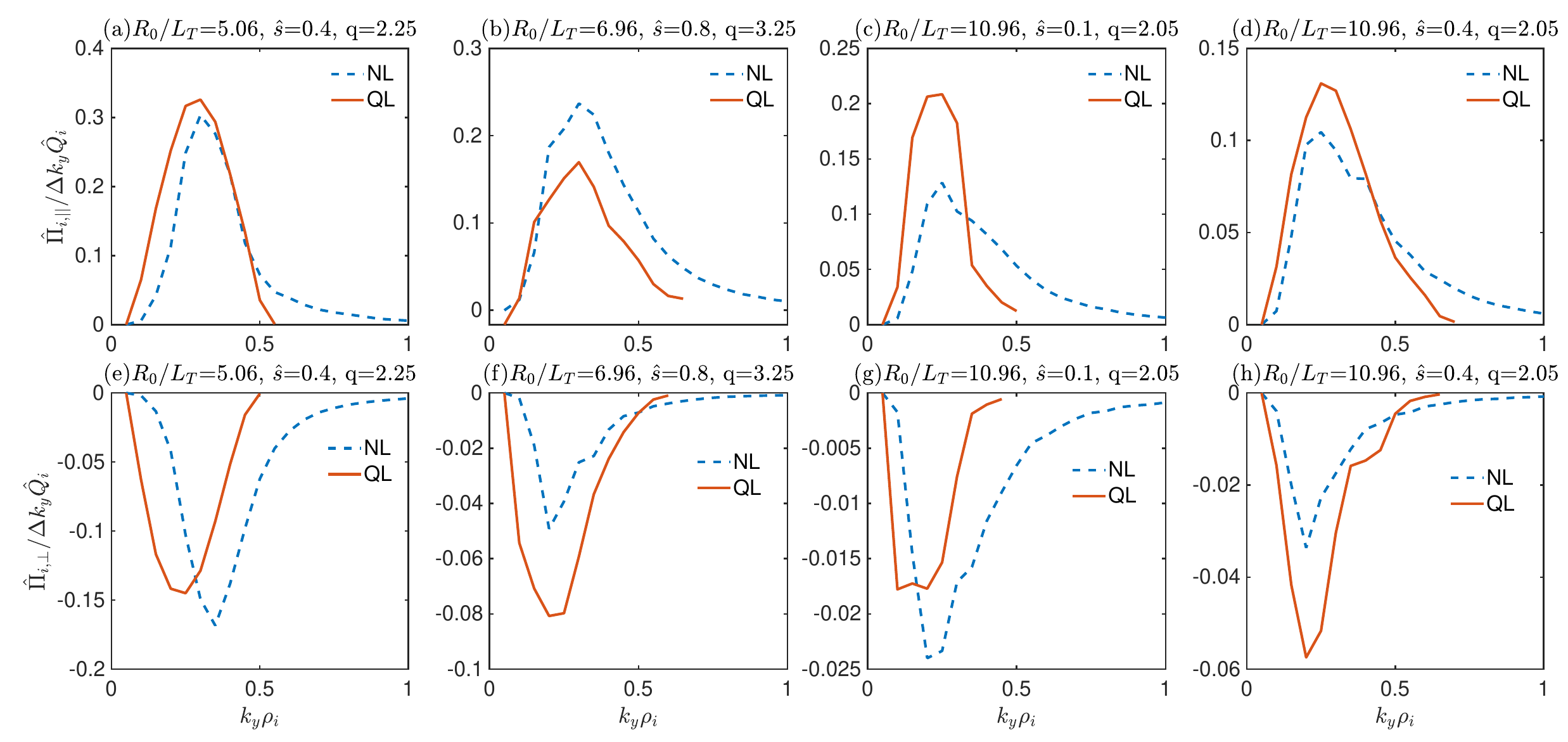}
\caption{\label{tightspectrum} A comparison of the $k_y$ spectra of the parallel $\hat{\Pi}_{i,||}$ (top) and perpendicular $\hat{\Pi}_{i,\perp}$ (bottom) components of the toroidal angular momentum flux obtained with NL simulations (dashed blue) and the flow shear QL model (solid red) for several representative tight aspect ratio cases with $\omega_{\perp}R_0/c_s=0.12$.}
%(a)-(d) compares the $k_y$ spectrum of $\Pi_{i,||}$ between NL simulations and QL estimates. (e)-(h) show the corresponding spectrum comparison for $\Pi_{i,\perp}$.
\end{figure}

\begin{figure}
\centering
\includegraphics[width=0.92\textwidth]{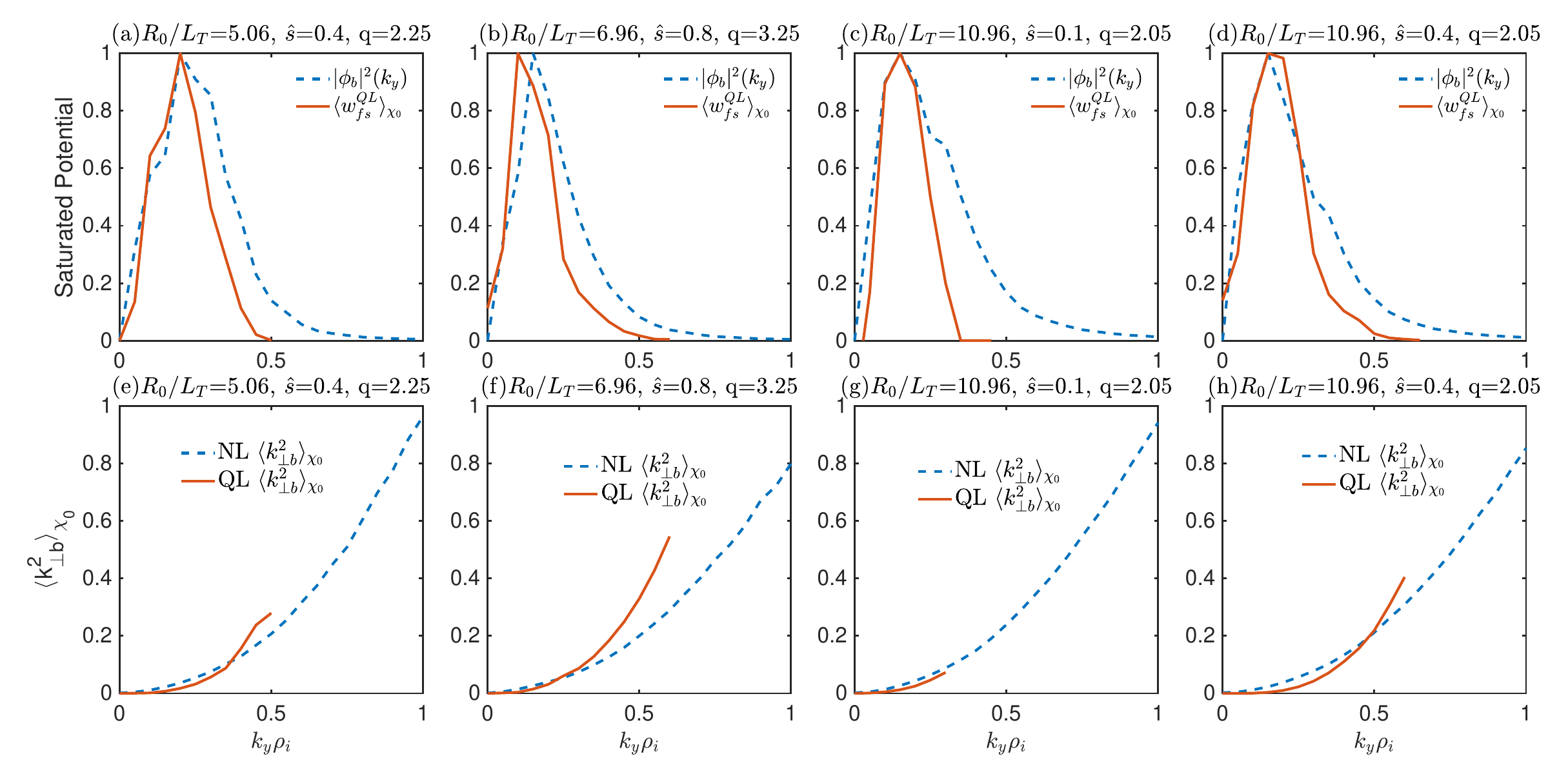}
\caption{\label{tightQLweightingkperp} A comparison of the ballooning angle averaged QL weights $\langle w^{QL}_{fs}\rangle_{\rchi_0}$ and perpendicular wave number $\langle k^2_{\perp b}\rangle_{\rchi_0}$ with NL simulations for several representative tight aspect ratio cases with flow shear, where $\langle...\rangle_{\rchi_0}$ denotes an average over all the ballooning angles in the simulations. (a)-(d) compare the $k_y$ spectra of the NL saturated potential averaged over all the ballooning angles $|\phi_b|^2(k_y)=\langle \text{MAX}_{z_b}\left[|\phi_b(\rchi_0,k_y,z_b,t)|^2\right]\rangle_{\rchi_0,t}$ with the QL estimate for this saturated potential given by $\langle w^{QL}_{fs}\rangle_{\rchi_0}$. Note that both the saturated potential and its QL estimate have been normalized to their maximum value since it is only the relative weighting of models that is important for the QL estimate of the Prandtl number. (e)-(h) show a comparison of $\langle k_{\perp b}^2\rangle_{\rchi_0}$ between NL simulations and the QL estimates.}
%(!!!The bracket might be added to gamma/kperp!!!)
\end{figure}

To carry out a more detailed comparison between NL and QL results, the $k_y$ spectra for $\hat{\Pi}_{i,||}$ and $\hat{\Pi}_{i,\perp}$ are shown in Fig. \ref{tightspectrum} for several different representative cases. Even the cases that are furthest from agreement in Fig. \ref{tightPrsum} display similar spectra between NL and QL results. For the same cases as Fig. \ref{tightspectrum}, Fig. \ref{tightQLweightingkperp} shows a detailed comparison between the QL weights and the NL square potential amplitudes $|\phi_b|^2(k_y)=\langle \text{MAX}_{z_b}\left[|\phi_b(\rchi_0,k_y,z_b,t)|^2\right]\rangle_{\rchi_0,t}$ (averaging over ballooning angle and time). As we know, the purpose of the QL weights is to estimate the relative values of the mode amplitude squared in the NL saturated state. As we can see in Fig. \ref{tightQLweightingkperp} (a)-(d), the $k_y$ spectral dependence of the NL potential matches well with the QL weight estimates. Figure \ref{tightQLweightingkperp} (e)-(h) shows the comparison of $\langle k^2_{\perp b}\rangle_{\rchi_0}$ between QL estimates and NL results, which also gives good agreement. This explains why our QL model generally gives accurate estimates. Importantly, we would not expect such a good agreement with other QL models, as they all, to the best of our knowledge, have used parallel momentum flux instead of toroidal angular momentum flux \cite{casson2009,Citrin2012QuaLiKizmagneticshear,Cottier2014QuaLiKizmomentum}. As shown by Fig. \ref{tightspectrum}, in the low Prandtl number regime (tight aspect ratio, low safety factor, and high magnetic shear), the perpendicular component $\hat{\Pi}_{i,\perp}$ of the toroidal angular momentum flux becomes large and cancels much of the parallel component $\hat{\Pi}_{i,||}$.

% \begin{figure}
% \centering
% \includegraphics[width=0.8\textwidth]{omt6.96shat0.8q3.25 spectrumreal2(method3-2).png}
% \includegraphics[width=0.8\textwidth]{omt6.96shat0.8q3.25 weightkperp(method3-2).png}
% \includegraphics[width=0.8\textwidth]{omt6.96shat0.8q3.25 flux 2Dspectrum(method3-2).png}
% \caption{\label{tight6.960.83.25spectrum} 1D spectrum comparison, QL weighting and $\langle k_{\perp}\rangle$ comparison, 2D spectrums of $\Pi_{||}$, $\Pi_{\perp}$, $Q$ and QL weighting given by our QL model as a function of $k_y$ and $\rchi_0$ for $R_0/L_T=6.96, \hat{s}=0.8, q=3.25$.}
% \end{figure}

% \begin{figure}
% \centering
% \includegraphics[width=0.8\textwidth]{omt5.06shat0.4q2.25 spectrumreal(method3-2).png}
% \includegraphics[width=0.8\textwidth]{omt5.06shat0.4q2.25 weightkperp(method3-2).png}
% \includegraphics[width=0.8\textwidth]{omt5.06shat0.4q2.25 flux 2Dspectrum(method3-2).png}
% \caption{\label{tight5.060.42.25spectrum} 1D spectrum comparison, QL weighting and $\langle k_{\perp}\rangle$ comparison, 2D spectrums of $\Pi_{||}$, $\Pi_{\perp}$, $Q$ and QL weighting given by our QL model as a function of $k_y$ and $\rchi_0$ for $R_0/L_T=5.06, \hat{s}=0.4, q=2.25$.}
% \end{figure}

\subsection{Model Benchmarking for more advanced cases}
\begin{figure}
\centering
\includegraphics[width=0.87\textwidth]{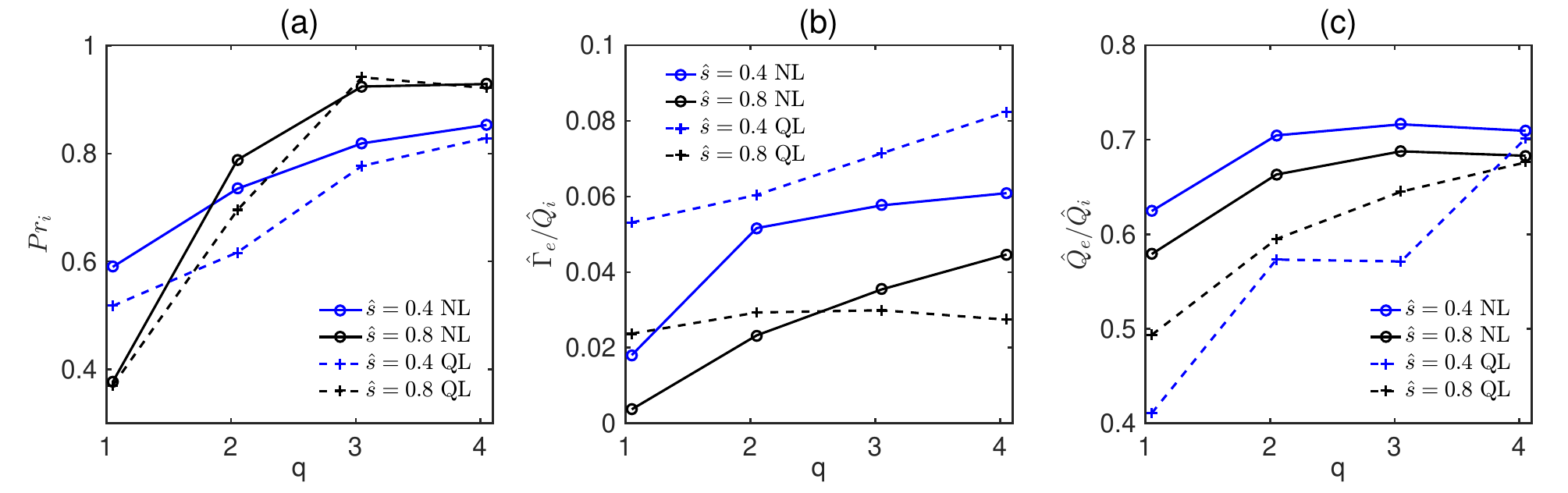}
\caption{\label{kineticPrsum} A comparison of the flux ratio between NL simulations (dashed) with kinetic electrons and our flow shear QL model (solid) for the (a) ion Prandtl number, (b) ratio of electron particle flux to ion heat flux, and (c) ratio of electron heat flux to ion heat flux.}
\end{figure}

In this section, we further benchmark our model for even more advanced cases to verify its general applicability. We first extended our tight aspect ratio cases with flow shear from the previous section by including fully kinetic electrons instead of adiabatic electrons. Table \ref{kineticphysicspara} shows the physical parameters used for these benchmark cases. The QL estimates for the Prandtl number are shown in Fig. \ref{kineticPrsum} (a), displaying a good match with the corresponding NL results. However, we can see that the match is somewhat better for higher magnetic shear than for lower magnetic shear. The average error for the $\hat{s}=0.8$ cases is only $15\%$, while the average error for the $\hat{s}=0.4$ cases is $20\%$. Figure \ref{kineticPrsum} (b) and (c) show that our QL model can also estimate other flux ratios such as $\hat{\Gamma}_e/\hat{Q}_i$ and $\hat{Q}_e/\hat{Q}_i$, where $\hat{\Gamma}_e$ is the particle flux $\Gamma_e$ normalized by $c_s n_i(\rho_i/R_0)^2$. Despite some mismatches (none of which exceed $30\%$), the agreement is good in general. This indicates that our QL model has the potential to be applied to estimate any flux ratios, not just the Prandtl number. 
%The $k_y$ spectrum comparison of several representative cases are further shown in Fig. \ref{kineticPrspectrum}.

\begin{table}\centering
\caption{\label{kineticphysicspara} The physical parameters used in GENE simulations of tight aspect ratio, circular geometries with kinetic electrons and flow shear. The grid parameters are the same as the previous adiabatic electron cases, which are shown in Tab. \ref{tightpara}.}
\footnotesize
\begin{tabular}{@{}llll}
\br
Parameter&Value\\
\mr\centering
Magnetic shear $\hat{s}$&$0.4,0.8$\\
Safety factor $q$&$1.05,2.05,3.05,4.05$\\
Inverse aspect ratio $\epsilon$&$0.36$\\
Temperature gradient $R_0/L_T$&$10.96$\\
Density gradient $R_0/L_n$&$2.22$\\
Flow shear $\omega_{\perp}R_0/c_s$&$0.12$\\ %=x_0/q_0\partial|\Omega_{tor}|/\partial x R_0/c_s
\br\centering
\end{tabular}\\
\end{table}
\normalsize

% \begin{figure}
% \centering
% % \includegraphics[width=0.95\textwidth]{omt10.96shat0.1q3.05kinetic1D spectrum.png}
% % \includegraphics[width=0.95\textwidth]{omt10.96shat0.4q3.05kinetic1D spectrum.png}
% % \includegraphics[width=0.95\textwidth]{omt10.96shat0.8q3.05kinetic1D spectrum.png}
% \includegraphics[width=0.92\textwidth]{Figure 15 kinetic spectrum.pdf}
% \caption{\label{kineticPrspectrum} Spectrum comparison between NL and QL estimate using our new model for several representative kinetic electron cases with tight aspect ratio and flow shear. (a)-(c) give the $k_y$ spectrum of $\Pi_{i,||}$ for NL and QL estimates using our new model. The cases are taken to be the ones with $q=2.05$, $\hat{s}=0.1,0.4,0.8$. (d)-(f) show the corresponding $k_y$ spectrum of $\Pi_{i,\perp}$.}
% \end{figure}

%The simulation parameters are shown directly in Fig. \ref{updownflowPrsum}.

In Sec. \ref{updownbenchmark} and Sec. \ref{tightadiabaticbenchmark}, we performed benchmarks for up-down asymmetric geometries without flow shear and for up-down symmetric geometries with flow shear, respectively. Here we consider several cases combining both the drive of momentum flux from up-down asymmetry and $\omega_{\perp}\ne 0$. Adiabatic electrons are again assumed. Such cases are of practical importance because they are needed to predict the actual rotation gradient that would arise in experiments. Specifically, the up-down asymmetry drives an intrinsic momentum flux, which we have calculated in Sec. \ref{updownbenchmark}. In an actual experiment, this intrinsic momentum flux will give rise to a rotation gradient that will quickly grow and drive a diffusive momentum flux. This diffusive momentum flux was calculated in Sec. \ref{tightadiabaticbenchmark}. In steady state and in the absence of external sources, these two fluxes must cancel \cite{ball2014,ball2016a,ball2018}. Otherwise, the finite momentum flux would cause the rotation profile to change in time. Thus, the rotation gradient expected in experiment is the one that achieves $\Pi_i=0$. Therefore, in order to determine the self-consistent effect of flow shear on the heat flux, one should scan the value of flow shear to find the value at which momentum flux $\Pi_i$ drops to zero and then look at the value of the heat flux. By doing so, we consistently determine how much flow shear will be self-generated as well as the corresponding steady state heat flux. In order to do this, it is important to efficiently find the value of flow shear $\omega_{\perp}$ that achieves $\Pi_i=0$ in up-down asymmetric geometries. This benchmark will show that our new flow shear QL model can achieve this. 

%, which we denote by $\omega_{\perp,0}$

\begin{figure}
\centering
\includegraphics[width=0.92\textwidth]{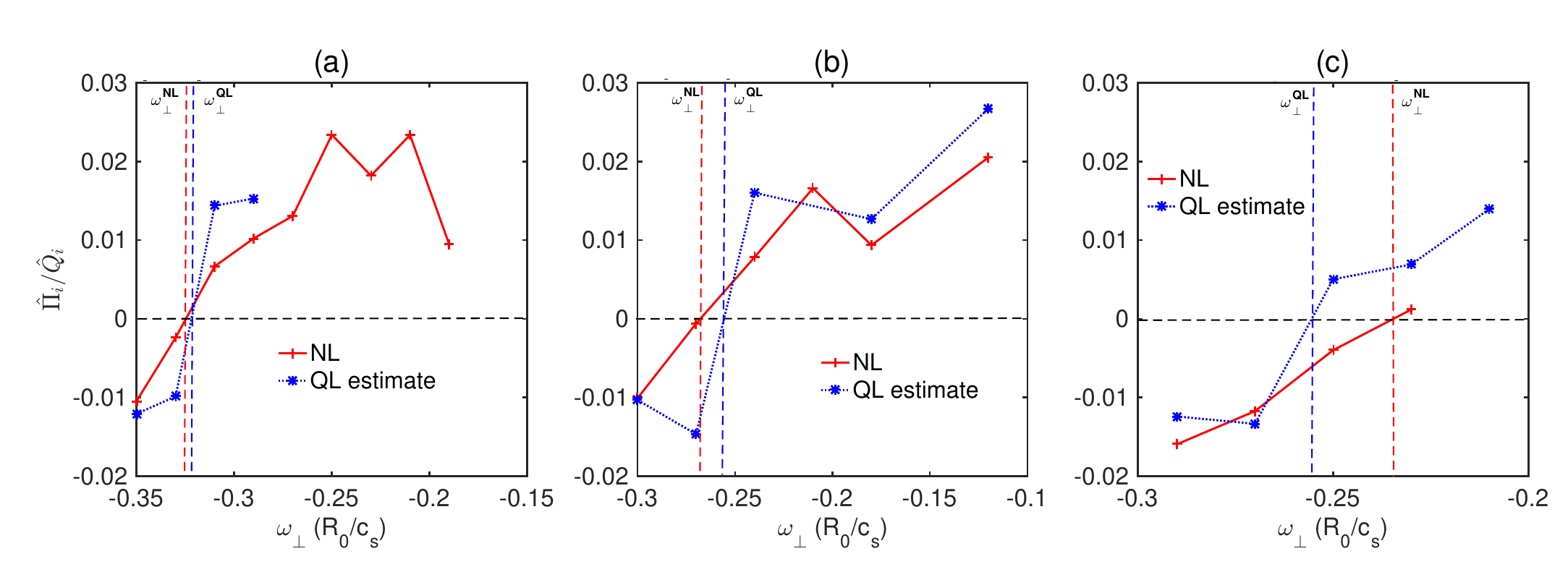}
\caption{\label{updownflowPrsum} A comparison of the flux ratio $\hat{\Pi}_i/\hat{Q}_i$ between NL simulations and the flow shear QL model for (a) $R_0/L_T=8.96$, (b) $R_0/L_T=10.96$ and (c) $R_0/L_T=12.96$. The dashed vertical lines indicate the value of $\omega_{\perp}$ that achieves $\Pi_i=0$.}
\end{figure}

\begin{figure}
\centering
\includegraphics[width=0.92\textwidth]{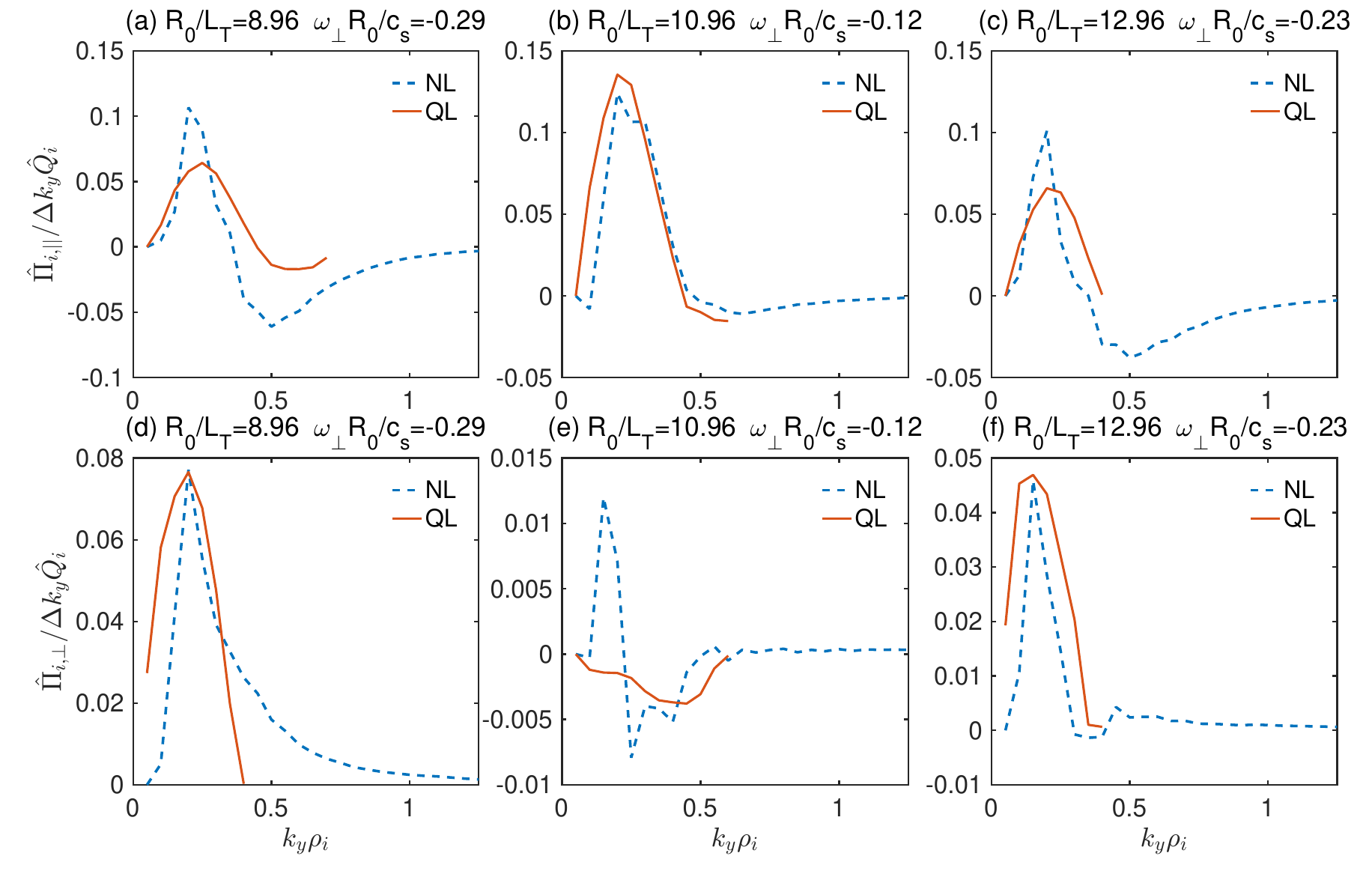}
\caption{\label{updownflowPrspectrum} A comparison of the $\hat{\Pi}_{i,||}$ (top) and $\hat{\Pi}_{i,\perp}$ (bottom) $k_y$ spectra between NL and the flow shear QL model for several representative up-down asymmetric equilibria with flow shear.}
%(a)-(c) give the $k_y$ spectrum of $\Pi_{i,||}$ for NL and QL estimates using our new model. (d)-(f) show the corresponding $k_y$ spectrum of $\Pi_{i,\perp}$.
\end{figure}

The simulation results are shown in Fig. \ref{updownflowPrsum}. As we can see, the QL model can provide good predictions of the flow shear value $\omega_{\perp}$ for which $\hat{\Pi}_i/\hat{Q}_i$ drops to zero. The average error in the zero crossing between QL (denoted by $\omega_{\perp}^{QL}$) and NL (denoted by $\omega_{\perp}^{NL}$) calculations is only $10\%$. Considering the fact that NL simulations have a statistical error and that the zero point is estimated by a linear interpolation, this is remarkably good agreement. We can also see that the value of $\hat{\Pi}_i/\hat{Q}_i$ is well-predicted by our flow shear QL model even away from the zero point. Figure \ref{updownflowPrspectrum} further shows the $k_y$ spectra of $\hat{\Pi}_{i,||}$ and $\hat{\Pi}_{i,\perp}$ for three representative cases. In general, these spectra also match fairly well with NL GENE simulations. Some deviation also occurs in sub-figure (a), (c) and (e), which indicates that combining flow shear and up-down asymmetry does make the QL estimate more challenging than the previous cases. Note that the flow shear values $\omega_{\perp}$ are negative. This is because the up-down asymmetric geometry we chose drives a positive intrinsic momentum flux (see Fig. \ref{Prupdownasy}). Thus, in order to cancel it, one must set $\omega_{\perp}$ to be negative in order to create a negative diffusive momentum flux.

% \begin{table}\centering
% \caption{\label{updownflowshearphysicspara}GENE physical parameters for tight aspect ratio $\epsilon=0.36$, adiabatic electrons, up-down asymmetric geometry simulation cases and flow shear with $\kappa=1.5$, $\theta_{\kappa}=\pi/8$, $R_0/L_n=2.22$, $\hat{s}=1.2$, $q=2.05$. The grid parameters are the same as previous adiabatic electron cases, see Tab. \ref{tightpara}.}
% \footnotesize
% \begin{tabular}{@{}cccc}
% \br
% Simulation Type &$\omega_{\perp}$&$R_0/L_T$\\%=x_0/q_0\partial|\Omega_{tor}|/\partial x R_0/c_s
% \mr\centering
% Nonlinear&$-0.35,-0.33,-0.31,-0.29,-0.27,-0.25,-0.23,-0.21,-0.19$&$8.96$\\
% Nonlinear&$-0.30,-0.27,-0.24,-0.18,-0.12$&$10.96$\\
% Nonlinear&$-0.29,-0.27,-0.25,-0.23,-0.21$&$12.96$\\
% \mr\centering
% Linear&$-0.35,-0.33,-0.31,-0.29$&$8.96$\\
% Linear&$-0.30,-0.27,-0.24,-0.21,-0.18,-0.12$&$10.96$\\
% Linear&$-0.29,-0.27,-0.25,-0.23$&$12.96$\\
% \br\centering
% \end{tabular}\\
% \end{table}
% \normalsize

\section{Conclusions and discussion}\label{section4}
%An alternative, but equivalent formulation is also presented, which avoids the need to trace a mode in time and reduces the computational cost by an additional factor of $3$.
In this paper, we constructed a new QL model to estimate momentum transport for micro-turbulence in the presence of rotational flow shear and strong up-down asymmetric plasma shaping. We first considered cases without flow shear but with up-down asymmetry to show the importance of considering multiple ballooning angles $\rchi_0$. Based on this observation, we extended the basic QL model to include multiple $\rchi_0$ values and showed the importance and validity of this approach. We then considered cases with flow shear, which required the construction of a new QL model to trace the time evolution of a ballooning mode as it moves across the $k_x$ domain. This flow shear QL model reduces to the previously obtained model when $\omega_{\perp}=0$ and has been thoroughly benchmarked, even for complex cases involving up-down asymmetry, flow shear, and low magnetic shear at tight aspect ratio. For the cases studied in this paper, the computational cost of using the most efficient full flow shear QL model is approximately $50$ times less than corresponding NL simulations. Although the Prandtl number estimates are somewhat less reliable near marginality, our model is shown to always give reasonable estimates for the flux ratio $\hat{\Pi}_i/\hat{Q}_i$. Our model is remarkably accurate at predicting the value of $\omega_{\perp}$ required to obtain a zero momentum flux in a simulation with up-down asymmetric geometry and flow shear, which is particularly relevant to experimental conditions. Future work will also account for other mechanisms driving momentum transport, most notably the pinch term.

%, based on the expression $Pr_i=\frac{\Pi_i}{Q_i}\frac{R_0}{L_T}\frac{\epsilon}{q\omega_{\perp}}$
Our QL model has a wide range of potential applications. First of all, it can be used to efficiently scope out parameter space to find the lowest Prandtl number before confirming these results with a small number of NL simulations. It can also be used in integrated modelling together with existing QL codes like TGLF. Given the $\text{Pr}_i\sim\hat{\Pi}_i/\hat{Q}_i$ from our model, $\hat{\Pi}_i$ and thus rotation profile can be estimated using the $\hat{Q}_i$ from TGLF. \textcolor{black}{Notice that in this approach, the accuracy of the estimate of $\hat{\Pi}_i$ will depend on both the accuracy of $\hat{Q}_i$ and the accuracy of the ratio $\hat{\Pi}_i/\hat{Q}_i$.} From an experimental perspective, our model can be used to quickly predict the Prandtl number in tokamaks, given an experimental measurement of flow shear $\omega_{\perp}$. Additionally, our model can estimate the amount of flow shear that would be driven by an external source of momentum (e.g., NBI). Similarly, for intrinsic rotation from up-down asymmetry, our model is able to estimate the self-consistent value of $\omega_{\perp}$ that will arise from a given geometry as has been done in this paper. 
%Given Tokamak à Configuration Variable (TCV) in Swiss Plasma Center has a great shaping capability, our future work will include applying our QL model to experimental predictions. 

%We noticed that this model is still not perfect. The Prandtl number estimates for several cases near marginality do not match the trend in NL GENE simulations. Our current model also only considered the momentum diffusivity but not other higher order global effects on momentum flux such as the pinch term, $\rho^*$ effects and neoclassical effects. A more comprehensive QL model for estimating global momentum transport will be our future investigations.

\clearpage
%%%%%%%%%%%%%%%%%%%%%%%%% Acknowledgement starts here!%%%%%%%%
\ack
The authors thank Antoine Cyril David Hoffmann, Alessandro Balestri and Oleg Krutkin for the fruitful discussions. This work has been carried out within the framework of the EUROfusion Consortium, via the Euratom Research and Training Programme (Grant Agreement No. 101052200 - EUROfusion) and funded by the Swiss State Secretariat for Education, Research and Innovation (SERI). Views and opinions expressed are however those of the authors only and do not necessarily reflect those of the European Union, the European Commission, or SERI. Neither the European Union nor the European Commission nor SERI can be held responsible for them.

\clearpage

\appendix

% \clearpage

\section{Expressions for the fluxes implemented in GENE code}\label{AppendixD}
This appendix gives the explicit expression of the particle flux, toroidal angular momentum flux, and heat flux implemented in GENE code. It is shown that the toroidal angular momentum flux has a non-trivial expression, which includes both a parallel component and a perpendicular component. Here we only take into account the electrostatic contribution to the turbulent fluxes. Thus, the explicit expressions for the fluxes in the Fourier space are \cite{Sugammafluxderiv1998,ParraUpDownSym2011,ball2016a}
%We will also briefly discuss the reason why the parallel component is usually diffusive while the perpendicular component is anti-diffusive. 
\begin{multline}\label{eq_explicitexpression_Gamma}
\Gamma_s(k_x,k_y,z,t)=-\frac{2\pi i}{m_s} C_y k_y B \phi(k_x,k_y,z,t)\int dv_{||}d\mu h_s(-k_x,-k_y,z,t)J_0(k_{\perp}\rho_s) ,
\end{multline}
% \be
\begin{multline}\label{eq_explicitexpression_Pi}
\Pi_s(k_x,k_y,z,t)=-\frac{2\pi i}{m_s} C_y k_y B \phi(k_x,k_y,z,t)\int dv_{||}d\mu h_s(-k_x,-k_y,z,t)\\
\times \left[\frac{I}{B}v_{||}J_0(k_{\perp}\rho_s)+\frac{i}{\Omega_s}\frac{dx}{d\psi}\frac{\mu k^x}{m_s}\frac{2J_1(k_{\perp}\rho_s)}{k_{\perp}\rho_s}\right],
\end{multline}
% \ee
and
\begin{multline}\label{eq_explicitexpression_Q}
Q_s(k_x,k_y,z,t)=-\frac{2\pi i}{m_s} C_y k_y B \phi(k_x,k_y,z,t)\int dv_{||}d\mu \frac{v^2}{2}h_s(-k_x,-k_y,z,t)J_0(k_{\perp}\rho_s).
\end{multline}
Here $h_s=\delta f_s+Z_se\phi F_{Ms}/T_s$ is the non-adiabatic portion of the distribution function, where $F_{Ms}$ is the background Maxwellian distribution, $Z_s$ is the charge number of the species being considered. $k^x=k_x|\Vec{\nabla}x|^2+k_y\Vec{\nabla}x\cdot\Vec{\nabla}y$, $\psi$ is the flux surface label, $C_y=(1/B_{ref}) d\psi/dx$ is a geometrical coefficient calculated by GENE based on flux surface information (where $B_{ref}$ is a reference value of the magnetic field), $\Omega_s=Z_seB/m_s$ is the particle gyro-frequency, $I=RB_{\zeta}$ is the toroidal field flux function and $J_0$ and $J_1$ are the zeroth and first order Bessel functions of the first kind. From Eq. \ref{eq_explicitexpression_Pi}, we can see that the toroidal angular momentum flux $\Pi_s=\Pi_{s,||}+\Pi_{s,\perp}$ can be distinguished into two contributions: the parallel component $\Pi_{s,||}$ and the perpendicular component $\Pi_{s,\perp}$, which are defined according to

\begin{multline}\label{eq_explicitexpression_Pipara}
\Pi_{s,||}(k_x,k_y,z,t)=-\frac{2\pi i}{m_s} C_y k_y B \phi(k_x,k_y,z,t)\int dv_{||}d\mu h_s(-k_x,-k_y,z,t)\frac{I}{B}v_{||}J_0(k_{\perp}\rho_s),
\end{multline}

\begin{multline}\label{eq_explicitexpression_Piperp}
\Pi_{s,\perp}(k_x,k_y,z,t)=-\frac{2\pi i}{m_s} C_y k_y B \phi(k_x,k_y,z,t)\int dv_{||}d\mu h_s(-k_x,-k_y,z,t)\\
\times\frac{i}{\Omega_s}\frac{dx}{d\psi}\frac{\mu k^x}{m_s}\frac{2J_1(k_{\perp}\rho_s)}{k_{\perp}\rho_s}.
\end{multline}

% \hfill \break
\clearpage

\section{Tables for grid parameters in benchmarks}\label{AppendixA}

%For each simulation, the number of independent ballooning modes (denoted by $M$) for NL, linear and linear $\rchi_0$ scan is $10\hat{s}k_y/k_{y,min}$, $1$, $1$, respectively. The $\rchi_0$ scan is over $[-\pi,\pi)$.

% \br\centering
% \multicolumn{9}{c}{Grid range}\\
% \mr\centering
% Simulation Type &$k_x\rho_i$&$k_y\rho_i$&$z$&$v_{||}/c_s$&$\sqrt{\mu/(T_i/B)}$&$t/(R_0/c_s)$\\
% \mr\centering
% NL&$[-0.1\pi k_{y,min}n_{k_x},0.1\pi k_{y,min}(n_{k_x}-1)]$&$[0.05,3.2]$&$[-\pi,\pi)$&$[-3,3]$&$(0,3)$&[0,1000]\\
% Linear&$[-\pi k_{y,min}\hat{s}n_{k_x},\pi k_{y,min}\hat{s}(n_{k_x}-1)]$&$[0.05,1]$ scan&$[-\pi,\pi)$&$[-3,3]$&$(0,3)$&[0,1000]\\
% Linear $\rchi_0$ scan&$[-\pi k_{y,min}\hat{s}n_{k_x}+\rchi_0 k_y\hat{s},\pi k_{y,min}\hat{s}(n_{k_x}-1)+\rchi_0 k_y\hat{s}]$, scan $\rchi_0$&$[0.05,1]$ scan&$[-\pi,\pi)$&$[-3,3]$&$(0,3)$&[0,1000]\\
% \mr\centering
% \multicolumn{9}{c}{Number of grid points}\\
% \mr\centering
% Simulation type &\multicolumn{3}{c}{$(n_{k_x},n_{k_y},n_z,n_{v_{||}},n_{\mu})$}&\multicolumn{3}{c}{$M$}\\
% \mr\centering
% NL&\multicolumn{3}{c}{$(192,64,32,32,9)$}&\multicolumn{3}{c}{$10\hat{s}k_y/k_{y,min}$}\\
% Linear&\multicolumn{3}{c}{$(192,1,32,32,9)$}&\multicolumn{3}{c}{$1$}\\
% Linear $\rchi_0$ scan&\multicolumn{3}{c}{$(12,1,32,32,9)$}&\multicolumn{3}{c}{$1$}\\
% \br\centering

% \begin{tabular}{@{}l}
% Grid Range\\
% \mr\centering
% \end{tabular}

% \multicolumn{9}{c}{Grid range}\\
% \mr\centering
\begin{table}[ht]%\left
\caption{\label{updownpara}The nominal GENE grid parameters for up-down asymmetric simulations with adiabatic electrons. The parameter $N_{\rchi_0}$ denotes the number of points in the $\rchi_0$ scan used in the simulation and $\Delta k_y\rho_i=0.05$. The physical parameters are shown in Tab. \ref{updownphysicspara}.}
\footnotesize
\scalebox{0.95}{
\begin{tabular}{@{}ccccccccc}%\hline
\br\centering
Simulation Type &$k_{x0}$&$\Delta k_x$&$N_{\rchi_0}$&$k_y\rho_i$&$z$&$v_{||}/\sqrt{2T_i/m_i}$&$\sqrt{\mu/(T_i/B)}$&$t/(R_0/c_s)$\\
\mr\centering
Nonlinear&$0$&$0.2\pi\Delta k_y\hat{s}$&1&$[0.05,3.2]$&$[-\pi,\pi)$&$[-3,3]$&$[0,3]$&[0,1000]\\
Linear&$0$&$2\pi k_y\hat{s}$&1&$[0.05,1]$ scan&$[-\pi,\pi)$&$[-3,3]$&$[0,3]$&[0,1000]\\
Linear $\rchi_0$ scan&$\rchi_0 k_y\hat{s}$&$2\pi k_y\hat{s}$&$\text{NINT}(10\hat{s})k_y/\Delta k_y$&$[0.05,1]$ scan&$[-\pi,\pi)$&$[-3,3]$&$[0,3]$&[0,1000]\\
\mr\centering
% \multicolumn{9}{c}{Number of grid points}\\
% \mr\centering
Simulation type &
\multicolumn{6}{c}{$(n_{k_x},n_{k_y},n_z,n_{v_{||}},n_{\mu})$}&
\multicolumn{2}{c}{$M$}\\
\mr\centering
Nonlinear &
\multicolumn{6}{c}{$(192,64,32,32,9)$}&\multicolumn{2}{c}{$\text{NINT}(10\hat{s})k_y/\Delta k_{y}$}\\
Linear &
\multicolumn{6}{c}{$(192,1,32,32,9)$}&\multicolumn{2}{c}{$1$}\\
Linear $\rchi_0$ scan &
\multicolumn{6}{c}{$(12,1,32,32,9)$}&\multicolumn{2}{c}{$1$}\\
\br\centering
\end{tabular}\\
}
\end{table}

% }
% \begin{tabular}{@{}ccc}\\
% \multicolumn{9}{c}{Grid Numbers}\\

% \end{tabular}
% \br\centering

% $^{a}$Inverse aspect ratio; \\
% $^{b}$Elongation of poloidal cross section; \\
% $^{c}$Tilt angle of the geometry. 

\begin{table}[ht]%\left
\caption{\label{tightpara} The nominal GENE grid parameters for tight aspect ratio ($\epsilon=0.36$) simulations with adiabatic electrons and flow shear $\omega_{\perp}R_0/c_s=0.12$, where $\Delta k_y\rho_i=0.05$. As we set $M$ greater than one, $k_{x0}$ is set to zero for all the cases in this table. The physical parameters are shown in Tab. \ref{tightphysicspara}.}
\footnotesize
\scalebox{0.849}{
\begin{tabular}{@{}ccccccccc}
\br\centering
% Simulation Type &$k_x\rho_i$&$k_y\rho_i$&$z$&$v_{||}/c_s$&$\sqrt{\mu/(T_i/B)}$&$t/(R_0/c_s)$&$(n_{k_x},n_{k_y},n_z,n_{v_{||}},n_{\mu},n_t)$&$M$\\
% \mr\centering
% Nonlinear&$[-0.1\pi k_{y,min}n_{k_x},0.1\pi k_{y,min}(n_{k_x}-1)]$&$[0.05,3.2]$&$[-\pi,\pi)$&$[-3,3]$&$(0,3)$&[0,1000]&$(192,64,32,32,9)$&$10\hat{s}k_y/k_{y,min}$\\
% Linear (normal $\hat{s}$)&$[-\pi k_{y,min}\hat{s}n_{k_x},\pi k_{y,min}\hat{s}(n_{k_x}-1)]$&$[0.05,1]$ scan&$[-\pi,\pi)$&$[-3,3]$&$(0,3)$&[0,1000]&$(192,1,32,32,9)$&$8$\\
% Linear (low $\hat{s}$)&$[-\pi k_{y,min}\hat{s}n_{k_x},\pi k_{y,min}\hat{s}(n_{k_x}-1)]$&$[0.05,1]$ scan&$[-\pi,\pi)$&$[-6,6]$&$(0,6)$&[0,1000]&$(768,1,32,64,18)$&$8$\\
Simulation Type &$\Delta k_x$&$k_y\rho_i$&$z$&$v_{||}/\sqrt{2T_i/m_i}$&$\sqrt{\mu/(T_i/B)}$&$t/(R_0/c_s)$&$(n_{k_x},n_{k_y},n_z,n_{v_{||}},n_{\mu},n_t)$&$M$\\
\mr\centering
Nonlinear&$0.2\pi\Delta k_y\hat{s}$&$[0.05,3.2]$&$[-\pi,\pi)$&$[-3,3]$&$[0,3]$&[0,1000]&$(192,64,32,32,9)$&$\text{NINT}(10\hat{s})k_y/\Delta k_{y}$\\
Linear (normal $\hat{s}$)&$0.25\pi k_y\hat{s}$&$[0.05,1]$ scan&$[-\pi,\pi)$&$[-3,3]$&$[0,3]$&[0,1000]&$(192,1,32,32,9)$&$8$\\
Linear (low $\hat{s}$)&$0.25\pi k_y\hat{s}$&$[0.05,1]$ scan&$[-\pi,\pi)$&$[-6,6]$&$[0,6]$&[0,1000]&$(768,1,32,64,18)$&$8$\\

\br\centering
\end{tabular}\\
}
\end{table}

% \clearpage
\hfill \break

\section{Calculating the flow shear QL model from a single time snapshot}\label{AppendixB}
%. However, since they are growing exponentially, we must initialize each of them properly to get their relative amplitudes correct. If this is done
The model constructed in Sec. \ref{section3} requires using time-dependent data from GENE linear simulations. When the magnetic shear is small, this requires a very high output frequency in order to fully resolve the remap time $\Delta t_{remap}=2\pi\hat{s}/M\omega_{\perp}$ \cite{Hammettposter}. Here we explain how the QL model with flow shear can be calculated from linear GENE data at a single time step, which is more computationally efficient. In practice, this is how we actually obtain the QL results in this paper. The key point is to set the initialization of the simulation very carefully. In the mode tracing method presented in Sec. \ref{section3}, we considered a single linear mode and, by following it in time, we could extract all the needed information for the flow shear QL model given by Eqs. \ref{eq_QL7-0} to \ref{eq_QL7-5}. To obtain the same result using data from a single snapshot in time will require us to consider all the different linear ballooning modes in the simulation (i.e., the different values of $\rchi_0^*$). In principle, all these linear modes should undergo the same evolution and thus carry the same information over time. However, this requires all these ballooning modes to be initialized in the same way. If these initializations of the modes for different $\rchi^{*}_{0}$ are correctly shifted in time, the state of the different linear modes at a given instant will exactly correspond to the history that you would find by tracing a single linear mode back in time. In traditional GENE simulations, the initial condition is to set all the Fourier modes in the system equal to a constant, which we will take to be $1$ (note that the absolute numerical values are not significant in linear results). Additionally, simulations with flow shear require a boundary condition for the $k_x$ grid: as the Fourier modes are pushed off on one side of the $k_x$ grid, they are simply discarded while the new modes that are added on the other side of the $k_x$ grid are initialized with zero amplitude. This approach, although reasonable, gives the ballooning modes with different $\rchi^{*}_{0}$ somewhat different initial conditions. Specifically, the linear modes that start on the grid do not have the same value as they are each at a different point in their Floquet period. 

\begin{figure}[ht]
\centering
\includegraphics[width=1.3\textwidth,height=0.8\textwidth]{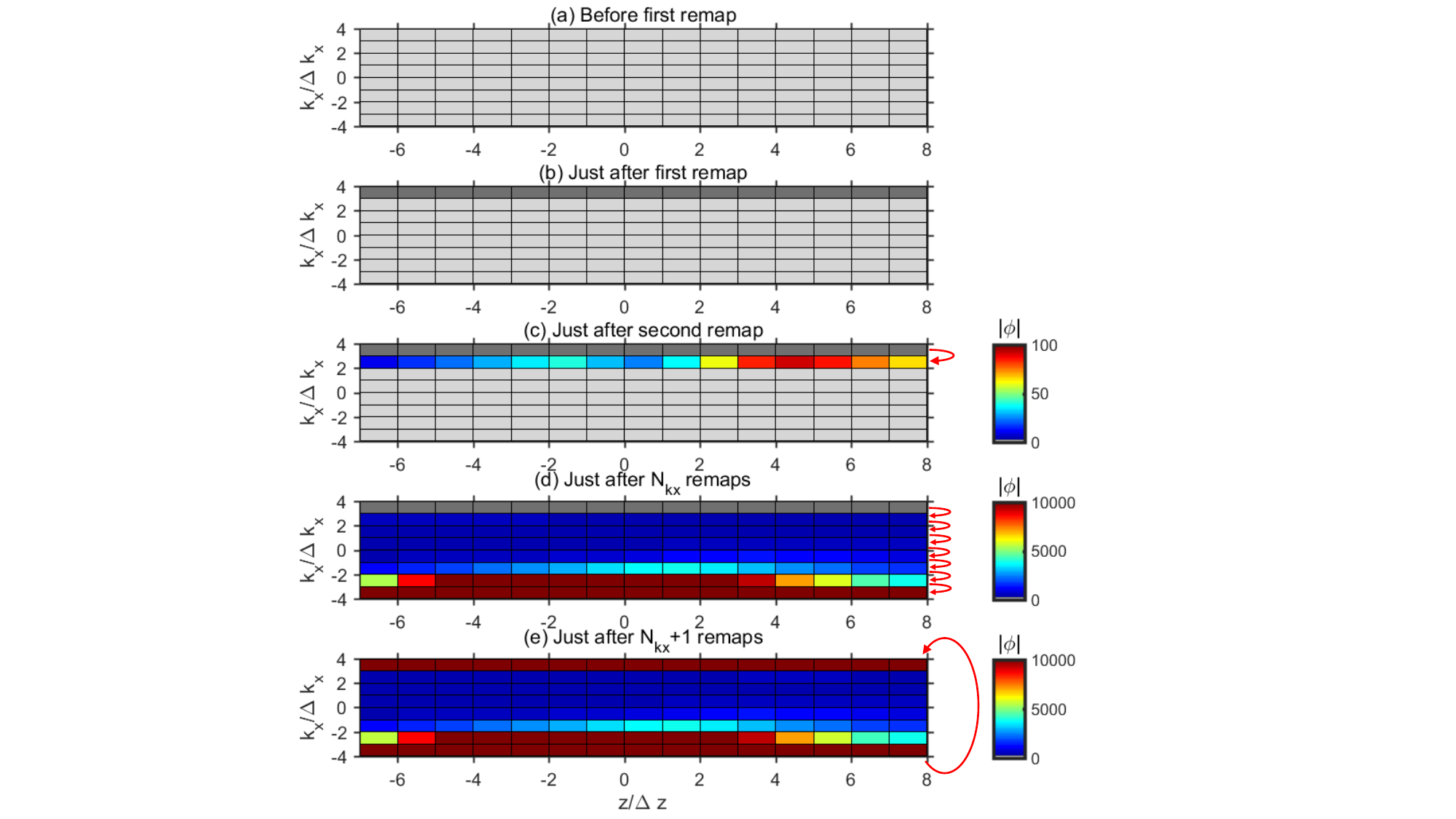}
\caption{\label{Remaptoy} A cartoon illustration of the initialization and boundary condition in $k_x$ used to calculate the flow shear QL model from a single snapshot in time. (a)-(e) show how the remap process work in the code. The dark grey color in (b), (c) and (d) indicates that the Fourier mode has a value of exactly $1$ (which is the case for a newly added mode), while the light grey color in (a), (b) and (c) denotes exactly $0$. $\Delta z$ in the label for the horizontal axis is $2\pi/n_z$. Please note that this figure represents just a toy simulation to illustrate the algorithm.}
\end{figure}

In our new approach, we initialize all of the Fourier modes on the grid to be zero, so nothing happens at first. Each time the flow shear remap occurs (at intervals $\Delta t_{remap}$), we feed Fourier modes with amplitude \say{$1$} into the system. After feeding $N_{kx}$ modes in, where $N_{kx}$ is the number of $k_x$ values on the grid, we switch to a periodic boundary condition, where the modes that fall off of the $k_x$ grid are immediately put back in on the other side. Figure \ref{Remaptoy} illustrates this process. In (a), before the first remap, the whole simulation domain is zero. When the first remap occurs, we must add a row of new values at either the maximum or minimum value of $k_x$ (depending on the sign of the flow shear $\omega_{\perp}$, here we choose the maximum value of $k_x$ because $\omega_{\perp}$ is positive). As is shown in (b), we set this new row to be $1$. In (c), the top most row is pushed downward to be the second row from the top and the top most line is again set to be $1$. In (d), this remap process has been repeated for $N_{kx}$ times (which is set to be $N_{kx}=8$ in this toy case) during which we keep on feeding ones into the top most row. Finally, in (e), when the non-zero data coming from the first remap is pushed off of the bottom row of the $k_x$ grid, we no longer feed ones, but instead the data being pushed out from the bottom row is moved to the top row. We then maintain this periodic boundary condition for the rest of the simulation. By doing so, the ballooning modes for different $\rchi^{*}_{0}$ are initialized in the same way. At a given time, the different ballooning structures for all $\rchi_0\in(-\pi,\pi]$ are representative of the evolution of a single ballooning structure over a full Floquet period. Therefore, every ballooning mode is identical and one can use different ballooning modes at a given time to get the same information as following a given mode back in time. To be specific, it means that $\phi_b(\rchi_0,k_y,z_b,t_{\infty}-\frac{\hat{s}}{\omega_{\perp}}(\rchi^*_0-\rchi_0))$ using the traditional initial condition is identical to $\phi_b(\rchi_0,k_y,z_b,t_{\infty})$ using our new initial condition. %As long as one takes the one last snapshot of the simulation, the 

This means that our QL model can be simplified in practice. If we use this special initialization, all the time tracing from Eq. \ref{eq_QL7-0} to Eq. \ref{eq_QL7-5} can be reduced to evaluations at $t_{\infty}$. The simplified formulas are
\be\label{eq_QLAp7-0}
F^{QL} = A_0\sum_{\rchi_{0},k_y}w_{fs}^{QL}(\rchi_{0},k_y)F^{L}_{norm}(\rchi_{0},k_y),
\ee

\be\label{eq_QLAp-1}
F^{L}_{norm}(\rchi_{0},k_y)=\frac{\left\langle F_{b}^{L}\left(\rchi_{0},k_y,z_b,t_{\infty}\right)\right\rangle_{z_b}}{\text{MAX}_{z_b}\left[\left|\phi_{b}\left(\rchi_{0},k_y,z_b,t_{\infty}\right)\right|^2\right]}, 
\ee

\be\label{eq_QLAp-2}
w_{fs}^{QL}(\rchi_{0},k_y)= 
\left\{
% \begin{matrix}
\begin{aligned}
%+\frac{\omega_{\perp}}{\hat{s}}t,
%\int_{0}^{\Delta\rchi_0(\rchi_0,k_y)} d\rchi_0^{'}\frac{\gamma( \rchi_0+\rchi_0^{'},k_y,t_{\infty})}{\langle k^2_{\perp b}\rangle(\rchi_0+\rchi_0^{'},k_y)}
\left(\frac{\Lambda}{\Delta \rchi_0(\rchi_0,k_y)}\right)^{\xi} \quad\quad\quad\quad\quad\quad\quad\quad\quad\quad\quad\quad\quad\quad\\       \quad\text{if}\quad \Lambda\equiv\int_{0}^{\Delta\rchi_0(\rchi_0,k_y)} d\rchi_0^{'}\frac{\gamma( \rchi_0-\rchi_0^{'},k_y,t_{\infty})}{\langle k^2_{\perp b}\rangle(\rchi_0-\rchi^{'}_0,k_y)}>0\\
0\quad \text{else},\quad\quad\quad\quad\quad\quad\quad\quad\quad\quad\quad\quad\quad\quad\quad\quad\quad \\
%\text{if} \int_{0}^{\Delta\rchi_0(k_y)} d\rchi_0\frac{\gamma( \rchi_0+d\rchi_0,k_y,t_{\infty}-\frac{\hat{s}}{\omega_{\perp}}(\rchi_{0}+d\rchi_0-\rchi^*_0))}{\langle k^2_{\perp}\rangle(\rchi_0+d\rchi_0,k_y)}<0,\xi=4
\end{aligned}
% \end{matrix}
\right.
\ee

\be\label{eq_QLAp-2-2}
\left\{
\begin{aligned}
\ln{\frac{|\phi_{b}(\rchi_{0},k_y,z_{b0},t_{\infty})|}{|\phi_{b}(\rchi_{0}-\rchi_{0,A_1}(\rchi_0,k_y),k_y,z_{b0},t_{\infty})|}}=A_1\ \\ 
A_1=O(1)\approx 1\quad\quad\quad\quad\quad\quad\quad\quad\quad\quad\quad\ \\
\Delta \rchi_0(\rchi_0,k_y)=\text{MIN}(\rchi_{0,A_1}(\rchi_0,k_y),2\pi),
\quad\quad\\
\end{aligned}
\right.
\ee

\be\label{eq_QLAp-3}
\gamma\left(\rchi_{0},k_y,t_{\infty}\right)=\frac{\omega_{\perp}}{\hat{s}\delta\rchi_0}\ln\left(\frac{|\phi_{b}(\rchi_{0},k_y,z_{b0},t_{\infty})|}{|\phi_{b}(\rchi_{0}-\delta\rchi_0,k_y,z_{b0},t_{\infty})|}\right),
\ee

\be\label{eq_QLAp-4}
\langle k_{\perp b}^2\rangle(\rchi_0,k_y)=\frac{\langle k_{\perp b}^2(\rchi_0,k_y,z_b)|\phi_{b}(\rchi_0,k_y,z_b,t_{\infty})|^2\rangle_{z_b}}{\langle |\phi_{b}(\rchi_0,k_y,z_b,t_{\infty})|^2\rangle_{z_b}},
\ee

\be\label{eq_QLAp-5}
\langle...\rangle_{z_b}=\frac{\int^{3\pi}_{-3\pi}dz_b(...)J(z_b)}{\int^{3\pi}_{-3\pi}dz_b J(z_b)}.
\ee

Therefore, when carrying out a QL estimate, we only need the last snapshot of the simulation. We mapped the time dependence of our model in Sec. \ref{modelreal} to the different ballooning angles. With this approach, as a result of a reduced number of output files, we found that the computational cost could be reduced by almost a factor of 3. The post-processing is also much more convenient because less output data is needed.

\section{Physics of mode advection in the ballooning space due to flow shear}\label{AppendixC}
\begin{figure}[ht]
\centering
\includegraphics[width=0.95\textwidth]{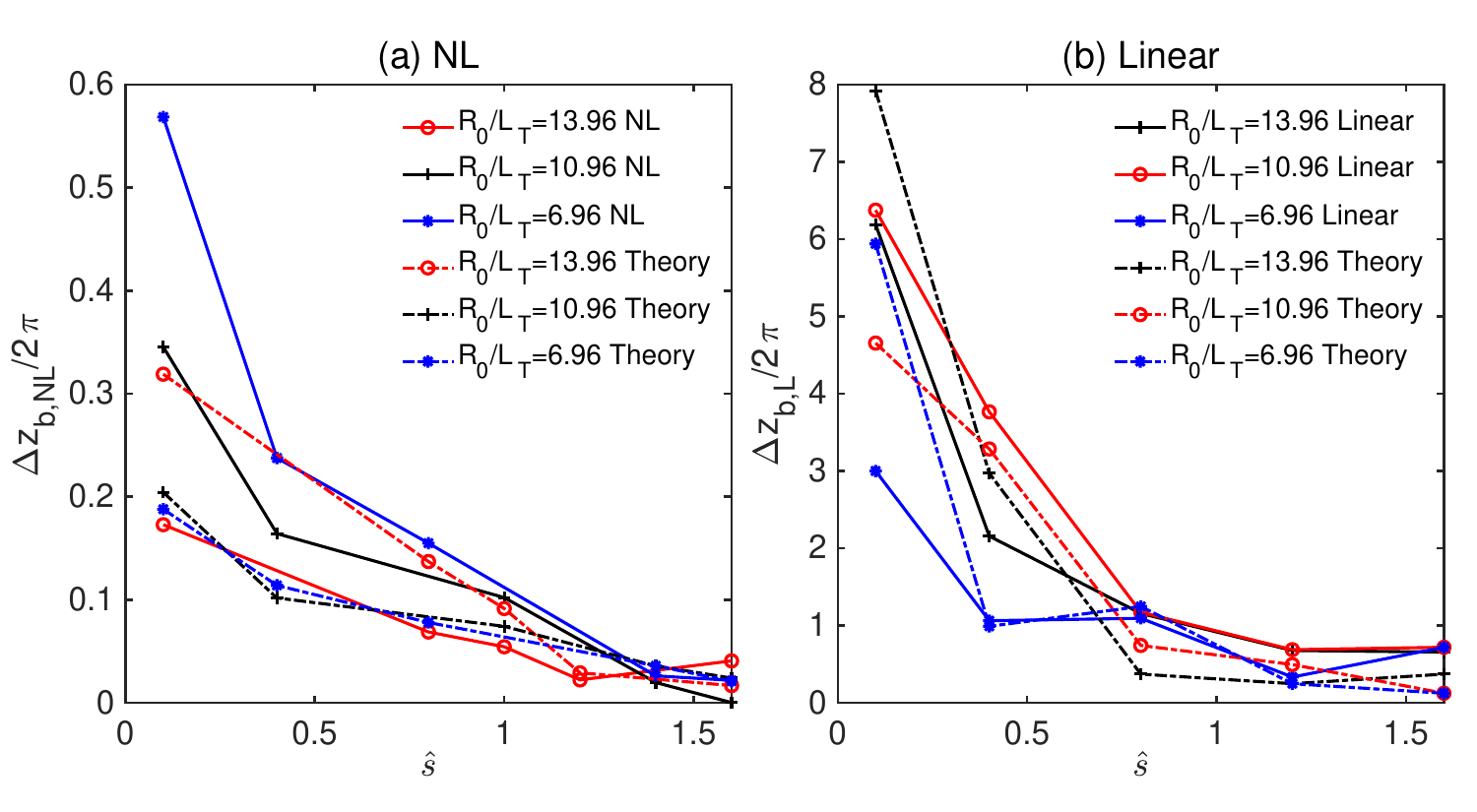}
\caption{\label{Tilt estimate} A comparison of the actual parallel location of the mode maximum (solid) and the theoretical estimate (dash dotted) with $\omega_{\perp}R_0/c_s=0.12$ for (a) NL simulations with $q=2.05$ and different $R_0/L_T$ and (b) linear simulations with $q=2.05$ and different $R_0/L_T$ as a function of $\hat{s}$. All the simulations use $k_y\rho_i=0.15$ mode and a circular geometry.}
\end{figure}

As shown in Fig. \ref{balltightshat0.10.8}, the NL ballooning structure remains around the central outboard midplane (i.e., $z_b\approx 0$) for both $\hat{s}=0.1$ and $\hat{s}=0.8$. However, in linear simulations, the ballooning structure is pushed far along the field line in $z_b$ when $\hat{s}=0.1$. In this Appendix, we will dive deeper into the physics behind this phenomenon and give a qualitative estimate of how far a mode moves due to the presence of flow shear. 
% \be\label{eq_QLApC-GK}
% \left(\frac{\partial}{\partial t}-\right),
% %\frac{\partial}{\partial t}\left(h-\frac{Ze\phi}{T}F_{M}\right)+\left(v_{||}+v_{M}\right)\cdot\nabla h+v_E\cdot\nabla\left(F_M+h\right)=0,
% \ee
First of all, we note that if there is no flow shear, the mode will peak in amplitude around $z_b=0$ in typical core simulations. Therefore, the \say{push} that moves the mode to larger $|z_b|$ must arise from perpendicular flow shear. In equilibrium, such a push by flow shear will be balanced by other effects that caused the mode peak at $z_b\approx 0$ in the absence of flow shear. Importantly, these can depend on if it is a linear simulation or a NL simulation, because the characteristic time scale in NL and linear simulation are different. To understand this, we can take a look at Eq. \ref{eq_floquet}, from which we can estimate how much a mode is shifted in the ballooning angle $\rchi_0$. At the same time, we know that $\rchi_0$ represents an estimate of the $z_b$ value at which the mode is aligned with $\Vec{\nabla} x$. Thus, they have a one-to-one correspondence, so we can estimate the mode shift in $z_b$ to be
%To understand this, we consider the gyrokinetic equation. Without collisions, the gyrokinetic equation with flow shear takes the form of Eq. (1) in \cite{ball2019}, where we can see that the perpendicular flow shear term (i.e. the one containing $\omega_{\perp}$) scales as $\omega_{\perp}k_y\frac{h}{\Delta k_x}$, where $h$ is the particle distribution function, $\Delta k_x=\rchi_{tilt}k_y\hat{s}$ is an estimate of how far the mode is shifted along the field line. In linear simulations, such a push is balanced by the linear mode growth term $\gamma h$, so we can write the following estimate
\be\label{eq_QLApC-1}
\Delta z_b\sim \Delta\rchi_0\sim \omega_{\perp}t/\hat{s}.
\ee
Using this expression and substituting the characteristic time scale for linear simulations $\tau_L\sim 1/\gamma$ (where $\gamma$ is the average linear growth rate for the fastest growing $k_y.$ mode in the presence of flow shear), we can deduce that the eigenfunction will peak at a distance of
\be\label{eq_QLApC-2}
\Delta z_{b,L} \sim\frac{1}{\gamma}\frac{\omega_{\perp}}{\hat{s}}
\ee
along the field line in ballooning space for linear simulations. In the NL simulations, the characteristic time scale becomes the NL decorrelation time $\tau_{NL}$, which measures the effect of the NL term. Therefore, the NL mode shift is estimated by
% \be\label{eq_NLApC-1}
% \frac{h}{\tau_{NL}}\sim\omega_{\perp}k_y\frac{h}{\Delta k_x},
% \ee
\be\label{eq_NLApC-2}
\Delta z_{b,NL}\sim\tau_{NL}\frac{\omega_{\perp}}{\hat{s}}.
\ee
In NL simulations, $\tau_{NL}$ can be estimated by the $\Delta t$ for which the correlation drops to $1/e$
\be\label{eq_NLApCdecorr}
C(\Delta t)=\langle\phi_{NZ}(x,y,z=0,t)\phi_{NZ}(x,y,z=0,t+\Delta t)\rangle_{x,y}/\langle|\phi_{NZ}|^2\rangle_{x,y},
\ee
where the subscript \say{NZ} denotes the non-zonal component. In this way, we obtain theoretical estimates for the location of the mode along the field line in both QL simulations (Eq. \ref{eq_QLApC-2}) and NL simulations (Eq. \ref{eq_NLApC-2}). These theoretical estimates can be compared with actual GENE simulations, as shown in Fig. \ref{Tilt estimate}. Note that there are no fitting parameters in the theoretical estimates, which indicates that the theory works well. 

As we see from the figure, the shift of the mode is almost always much weaker nonlinearly than linearly. Additionally, the shift typically increases with the decrease of $\hat{s}$ as predicted by our theoretical estimates. Comparing the lines with the same color in the figure, we see a quite good match. The agreement is a bit worse at low $\hat{s}$, but we see the theory captures the most important trend, i.e., the factor of $10$ difference (comparing Fig. \ref{Tilt estimate} (a) and Fig. \ref{Tilt estimate} (b)) between linear and NL simulations. This shows that our estimates given by Eq. \ref{eq_QLApC-2} and Eq. \ref{eq_NLApC-2} are reasonable. 

Note that large shifts make QL estimates more challenging. For our flow shear QL model, the integration in ballooning space is taken from $-3\pi$ to $3\pi$ (see Eq. \ref{eq_QL7-5} and Eq. \ref{eq_QLAp-5}) for the reasons explained in Sec. \ref{ballspacecomparesec}. Combining this information with Eq. \ref{eq_QLApC-2} by setting $\Delta z_{b,NL}=3\pi$, we found that our QL model will work if the following condition is satisfied
\be\label{eq_criteria}
\frac{\omega_{\perp}}{\hat{s}}<3\pi\frac{1}{\tau_{NL}} \approx 4,
\ee
which gives a criteria for the validity of our QL model. %In principle, one could change the range of integration to make the QL model valid for different magnetic shear values.

% \clearpage

\section*{References}
\bibliography{sample}
\end{document}